\documentclass[preprint,12pt]{elsarticle}
\usepackage{graphicx}
\usepackage{subcaption}  % Use subcaption instead of subfigure
\usepackage{float}
\usepackage{caption}  % Better caption control

\usepackage{amssymb}
\usepackage{amsmath}
\usepackage[thmmarks, amsmath, amsthm]{ntheorem}

\newtheorem{theorem}{Theorem}[section]

\newtheorem{problem}{Problem}
\newtheorem{definition}{Definition}
\newtheorem{assumption}{Assumption}
\newtheorem{remark}{Remark}

\newcommand{\Qd}{\dot{Q}}
\newcommand{\Qdi}[1]{\dot{Q}_i^{#1}}

\newcommand{\pd}{\dot{p}}

\newcommand{\Eti}{{E_{t,i}}}

\newcommand{\zd}{\dot{z}}
\newcommand{\sign}{\text{sign}}
\newcommand{\Pref}[1]{\ref{#1}}

\journal{Electric Power Systems Research}

\begin{document}

\begin{frontmatter}

\title{Distributed component-level modeling and control of energy dynamics in electric power systems} 

\author[MIT]{Hiya Gada}\ead{hiyagada@mit.edu}
\author[SmartGridz]{Rupamathi Jaddivada}\ead{jrupamathi93@gmail.com}
\author[MIT]{Marija Ilic\corref{cor1}}\ead{ilic@mit.edu}  

\cortext[cor1]{Corresponding author. Tel. +1-617-324-0645}

\affiliation[MIT]{organization={Laboratory for Information and Decision Systems, Massachusetts Institute of Technology},
            % addressline={}, 
            city={Cambridge},
            postcode={02139}, 
            state={Massachusetts},
            country={USA}}
\affiliation[SmartGridz]{
            organization={SmartGridz},
            % addressline={}, 
            city={Sudbury},
            postcode={01776}, 
            state={Massachusetts},
            country={USA}}

\begin{abstract}
The widespread deployment of power electronic technologies is transforming modern power systems into fast, nonlinear, and heterogeneous networks.
Conventional modeling and control approaches, rooted in quasi-static analysis and centralized architectures, are inadequate for these converter-dominated systems operating on fast timescales with diverse and proprietary component models.
% This paper adopts and extends the energy space modeling framework introduced by Ilic and Jaddivada (2018, 2021) and grounded in energy conservation principles to address these challenges.
This paper adopts and extends a previously introduced energy space modeling framework grounded in energy conservation principles to address these challenges.
We generalize the notion of a \textit{port interaction variable}, which encodes energy exchange between interconnected components in a unified manner.
A multilayered distributed control architecture is proposed in which dynamics of each component are lifted to a linear energy space through well-defined mappings.
Distributed control with provable convergence guarantees is derived in energy space using only local states and minimal neighbor information communicated through port interactions.
The framework is validated using two examples: voltage regulation in an inverter-controlled RLC circuit and frequency regulation of a synchronous generator. 
The energy-based controllers show improved transient and steady-state performance with reduced control effort compared to conventional methods.
\end{abstract}

\end{frontmatter}
\section{Introduction}
\label{sec:introduction}
% Fast
Modern electric power systems are rapidly evolving with the integration of power electronic technologies, including inverter-based resources, battery energy storage systems, high-voltage direct current links, and flexible alternating current transmission systems~\cite{milano2018foundations,fang2018inertia,crivellaro2020beyond}.
Traditional modeling and control frameworks rely on assumptions, such as quasi-static phasor representations and PQ decoupling~\cite{sauer1998power}, that break down in systems dominated by fast, nonlinear, and tightly coupled converter dynamics.
% Nonlinear
Moreover, most components in modern power systems exhibit inherently nonlinear dynamics~\cite{darbali2019nonlinear,nugroho2019nonlinearity}.
While nonlinear feedback control can regulate an isolated component with known dynamics~\cite{he2022nonlinear}, extending such guarantees to interconnected systems with dynamic cross-coupling and limited model transparency is nontrivial~\cite{venkatramanan2022integrated,johnson2022generic}.
% Interconnections introduce dynamic cross-couplings with neighboring subsystems whose dynamics may be unknown, preventing direct application of standard single-system stability analyses~\cite{venkatramanan2022integrated,johnson2022generic}.

% Heterogeneity and distributed control
These challenges are compounded by growing grid heterogeneity. 
Components developed by different vendors rely on proprietary models and control architectures, rendering centralized coordination impractical \cite{johnson2022generic}.
This necessitates abstractions that can unify component dynamics in a modular manner~\cite{venkatramanan2022integrated,markovic2021understanding}.
Distributed control offers a scalable alternative by enabling local decisions based on local measurements and limited peer-to-peer communication, while achieving global objectives such as frequency regulation~\cite{wang2024distributed}.

% Relation to Existing Energy-Based Modeling and Control Frameworks
Energy-based modeling and control provide a natural foundation for addressing nonlinear interconnected systems. 
Dissipativity theory, based on supply and storage functions, establishes dissipation inequalities from energy conservation laws and has been widely used for passivity-based control~\cite{willems1972dissipative,hill1976stability}.
However, classical passivity formulations impose hard constraints on controller energy extraction, often referred to as pervasive dissipation.
Power-shaping techniques were later introduced to mitigate this limitation by using passivity properties with differentiated port variables~\cite{garcia2004power,ortega2003power}, however, their applicability does not extend straightforwardly to general RLC networks and other mechanical systems. 
Brayton–Moser modeling \cite{brayton1964nonlinear} introduced mixed potential functions to describe RLC networks and later mechanical analogues. 
Although structurally insightful, most Brayton–Moser approaches require complete network descriptions or centralized reformulation procedures that are not inherently modular or distributed~\cite{jeltsema2003modeling}.

% Energy space motivation
Motivated by the need for a modeling framework compatible with distributed architectures,
this paper builds upon prior work in energy space modeling~\cite{ilic2018multi,ilic2018fundamental,ilic2020unified}, grounded in energy conservation and system interconnection principles. 
Unlike traditional state-space models based on voltage and current variables, the energy space formulation adopts energy as the state variable, resulting in technology-agnostic and linear dynamics. 
Thus, it is well suited to capture fast transients in converter-dominated systems while providing a unified and modular structure that naturally aligns with distributed control architectures~\cite{cvetkovic2015interaction}.

% Interaction variable
A central construct within energy space modeling is the notion of a \emph{port interaction variable}, which captures energy exchange between subsystems via shared ports.
These variables are defined in terms of power flows, and serve as the communicated quantities from neighbors in distributed control.
Over the last two decades, the interaction variable-based theory has been utilized to design nonlinear controllers for inverters, induction machines, generators, and residential demand-side technologies~\cite{miao2020high,ilic2020plug,ilic2019toward}.
This work builds on these foundations to formulate a distributed control problem in energy space and develop controllers compatible with fast, nonlinear, and heterogeneous component dynamics.

% Outline of paper
Section~\ref{sec:distributed_general} formulates a general distributed control problem for a power system modeled as interconnected subsystems with local dynamics subject to port, control, and exogenous inputs. 
Section~\ref{sec:energy_modeling} reviews the energy space modeling framework used throughout the paper. 
We summarize preliminaries~\cite{ilic2018multi,ilic2018fundamental,ilic2020unified} and introduce an extended third-order energy space model that captures the dynamics of stored energy in tangent space, previously treated as an independent disturbance. 
This extension allows accurate modeling of fast-timescale behavior. 
Section~\ref{sec:distributed_energy} establishes a multilayered architecture in which each component’s physical state is lifted to energy space, an energy space control input is designed using local and communicated neighbor information, and the resulting control is mapped back to the physical space for implementation. 
While control objectives in interconnected systems may include system-level coordination of power exchanges, this work focuses on component-level stability, assuming that system-level interaction references are provided.
In Section~\ref{sec:energy-based_component-level_control}, standard nonlinear control techniques, namely feedback linearization and sliding mode control, are adapted to energy space, and sufficient conditions for asymptotic output error convergence are established.
Section~\ref{sec:examples} illustrates the proposed architecture using an RLC circuit and a synchronous machine, with voltage and frequency regulation as their respective objectives. 
The energy-based controllers show improved transient and steady-state performance with reduced control effort compared to conventional methods.
Section~\ref{sec:conclusion} presents concluding remarks and directions for future work.

\section{Distributed control problem formulation}
\label{sec:distributed_general}

\begin{figure}[ht]
\begin{center}
\includegraphics[width=0.75\linewidth]{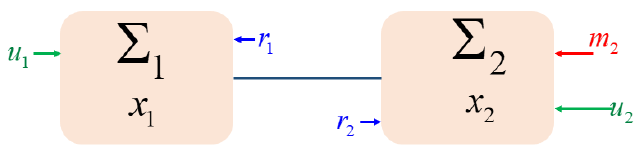}
\caption{Interconnected system comprising two components $\Sigma_1$ and $\Sigma_2$ with local controllable inputs $u_1, u_2$, port inputs $r_1, r_2$, and exogenous disturbance $m_2$}
\label{fig:system_general}
\end{center}
\end{figure}

To formally pose the distributed control problem, we review the modeling approach presented in~\cite{willems2007behavioural}, which represents the system as an interconnection of components, as illustrated in Figure~\ref{fig:system_general}.
Let $x_i \in \mathcal{X}_i$ denote the state variable of component $\Sigma_i$, capturing its internal dynamics, and let $u_i \in \mathcal{U}_i$ denote the locally implemented control input. 
The external inputs $r_i \in \mathcal{R}_i$ and $m_i \in \mathcal{M}_i$ denote, respectively, the port input that encodes the interaction with the rest of the system, and the exogenous disturbances.

The dynamic model of component $\Sigma_i$ is described by,
% Without loss of generality, we assume that $u_i$, $r_i$, and $m_i$ enter the state dynamics affinely through state-dependent input functions.
\begin{subequations}
\label{Eqn:GeneralCompModel_COnv}
%%%%
\begin{align}
&\text{Initial condition:} \quad x_i(0) = x_{i,0},  \\
&\text{State dynamics:} \quad
 \dot{x}_i = f_{x,i}(x_i) + g_i^r(x_i)\, r_i + g_i^m(x_i)\, m_i + g_i^u(x_i)\, u_i,
\label{Eqn:StateGeneralDyn} \\
&\text{Output of interest:} \quad y_i = f_{y,i}(x_i, u_i, r_i, m_i), \\
&\text{Common output:} \quad z_i = f_{z,i}(x_i, u_i, r_i, m_i).
\end{align}
\end{subequations}
Here, $y_i$ represents the component’s output of interest, such as frequency or voltage, while $z_i$ represents the common output that is physically shared between neighboring components through interconnection, such as current at a point of common coupling or exchanged power.
To ensure feasibility, $z_i$ must satisfy the interconnection condition~\eqref{eqn:interconnection_condition_general} imposed by conservation laws, (e.g., Tellegen's theorem \cite{tellegen}), making it a suitable choice of information to be communicated between components for distributed control. 

A key difficulty in distributed control is relating control specifications on $y_i$ and the common output $z_j$ communicated by neighboring components.
Assuming a map $\psi_i$ of the form~\eqref{eqn:yref_dist} exists, the output reference is defined as,
\begin{align} \label{eqn:yref_dist}
    y^{ref}_i = \psi_i (x_i, \{z_j\}_{j \in \mathcal{C}_i}),
\end{align}
where $\mathcal{C}_i$ denotes the set of components that are directly connected to $\Sigma_i$ (i.e., its neighbors). 
This formulation ensures consistency between local control objectives and network-level interactions.

% The distributed control problem is formally posed in Problem \Pref{prob:control_objectives}.

\begin{problem}[Distributed control problem] \label{prob:control_objectives}
    Given a system consisting of a set $\mathcal{N}$ of interconnected subsystems, the goal is to design a smooth feedback control law of the form,
    \begin{align*}
        u_i = c_i(x_i, y^{ref}_i, \{z_j\}_{j \in \mathcal{C}_i}) , \quad \forall i \in \mathcal{N},
    \end{align*}
    that utilizes local state information $x_i$ and minimal information $z_j$ communicated by neighboring components $\Sigma_j$, such that the following is satisfied:
    \begin{enumerate}
        \item \textbf{Interconnection constraint:}
        For each interconnection, the common output variables of neighboring components must satisfy,
        \begin{align} \label{eqn:interconnection_condition_general}
            z_i = - \sum_{j\in \mathcal{C}_i} z_j, \quad \forall i \in \mathcal{N}.
        \end{align}
        \item \textbf{Control objective:}
         The local output of interest $y_i$ must converge to its reference value given by the map $\psi_i$ as defined in \eqref{eqn:yref_dist}, i.e.,
        \begin{align*}
            y_i \to y_i^{ref} \quad \text{as  } t \to \infty, \quad \forall i \in \mathcal{N}.
        \end{align*}
    \end{enumerate}
\end{problem}

\section{Energy-based modeling of a component}
\label{sec:energy_modeling}
To provide a physically intuitive, unified representation across heterogeneous technologies, we map the component's conventional state dynamics to linear {\textit{energy space dynamics}}.

\subsection{Preliminaries}
We review definitions from the energy-based modeling framework in~\cite{ilic2018multi, rupa2021feasible} and extend it to capture higher-order energy dynamics.
\begin{definition} [Stored energy] \label{Defn:StoredEnergy}
The stored energy of component $\Sigma_i$ is described by the energy function $E_i~:~\mathcal{X}_i~\to~\mathbb{R}$, defined as,
\begin{align*}
E_i(x_i) = \frac{1}{2} x_i^\top H_i(x_i) x_i,
\end{align*}
where $H_i(x_i)$ is a positive definite symmetric inertia matrix for all $x_i \in \mathcal{X}_i$.
\end{definition}

The inertia matrix $H_i(x_i)$ represents inductance/capacitance in electrical systems and mass/inertia in mechanical systems~\cite{ilic2018multi,Jeltsema2009}.
\begin{definition} [Stored energy in tangent space]\label{Defn:StoredEnergyTangent}
For the same $H_i$ as in Definition~\ref{Defn:StoredEnergy}, stored energy in tangent space $E_{t,i}$ is defined over the tangent bundle $\mathcal{T} \mathcal{X}_i$, where $(x_i, \dot{x}_i) \in \mathcal{T} \mathcal{X}_i$, through the mapping $E_{t,i} : \mathcal{T} \mathcal{X}_i \to \mathbb{R}$ as,
\begin{align*}
E_{t,i}(x_i, \dot{x}_i) = \frac{1}{2} \dot{x}_i^\top H_i(x_i) \dot{x}_i.
\end{align*}
\end{definition}

While not directly associated with physical storage elements, the tangent-space energy extends the energy state formulation in~\cite{ilic2018multi, rupa2021feasible} to enable a linear representation of higher-order energy dynamics (see~\eqref{eq:en_state_model}). 
\begin{definition} [Dissipation]\label{Defn:Dissip}
The instantaneous power dissipated by the component $\Sigma_i$ is described by a dissipation function $D_i : \mathcal{X}_i \to \mathbb{R}$, defined as,
\begin{align*}
D_i(x_i) = \frac{1}{2} x_i^\top B_i(x_i) x_i,
\end{align*}
where $B_i(x_i)$ is a positive semi-definite matrix for all $x_i \in \mathcal{X}_i$.
\end{definition}
\begin{definition} [Dissipation in tangent space]\label{Defn:DissipTangent}
For the same $B_i(x_i)$ as in Definition~\ref{Defn:Dissip}, the instantaneous power dissipated in tangent space by $\Sigma_i$ is,
\begin{align*}
D_{t,i}(x_i, \dot{x}_i) = \frac{1}{2} \dot{x}_i^\top B_i(x_i) \dot{x}_i.
\end{align*}
\end{definition}
\begin{definition}[Time constants] \label{defn:TimeConstants}
The time constant and tangent-space time constant of $\Sigma_i$ are defined as,
\begin{align*}
\tau_i(x_i) = \frac{E_i(x_i)}{D_i(x_i)}, \quad
\tau_{t,i}(x_i, \dot{x}_i) = \frac{E_{t,i}(x_i, \dot{x}_i)}{D_{t,i}(x_i, \dot{x}_i)},
\end{align*}
provided that $D_i(x_i) > 0$ and $D_{t,i}(x_i, \dot{x}_i) > 0$.
\end{definition}
We now formalize port, control, and disturbance inputs in energy space as counterparts of the conventional inputs in Section~\ref{sec:distributed_general}, using effort and flow variables. 
Let $\mathcal{E}_i$ and $\mathcal{F}_i$ be dual spaces with $\mathcal{E}_i = (\mathcal{F}_i)^*$ and $\mathcal{F}_i = (\mathcal{E}_i)^*$.
Elements $e_i \in \mathcal{E}_i$ and $f_i \in \mathcal{F}_i$ denote effort and flow variables~\cite{WyattIlic1990}, representing generalized forces (e.g., voltage, mechanical force) and their conjugates (e.g., current, velocity), respectively.
Using these variables, we define instantaneous power, rate of change of generalized reactive power, and tangent-space power associated with control, port, and exogenous inputs.
\begin{definition}[Instantaneous power]\label{Defn:RealPower}
For input $\alpha \in \{r,u,m\}$ (i.e., port, control, disturbance),
the instantaneous power into a component $\Sigma_i$ is,
\begin{align*}
    P_i^\alpha =  e_i^\alpha f_i^\alpha,
\end{align*}
where $e_i^\alpha$ and $f_i^\alpha$ are the effort and flow variables associated with input $\alpha$.
\end{definition}
We review the definition of the rate of change of generalized reactive power introduced in~\cite{ilic2018fundamental}, and restate it here to clarify its structure within the energy-based modeling framework. 
\begin{definition} [Rate of change of generalized reactive power]\label{Defn:ReacPower}
For input $\alpha \in \{r,u,m\}$ (i.e., port, control, disturbance), the rate of change of generalized reactive power into a component $\Sigma_i$ is,
\begin{align*}
    \dot{Q}_i^\alpha = e_i^\alpha \frac{d f_i^\alpha}{dt} - f_i^\alpha \frac{d e_i^\alpha}{dt}
\end{align*}
where $(e_i^\alpha, \dot{e}_i^\alpha)  \in \mathcal{T}\mathcal{E}_i^\alpha$ and $(f_i^\alpha, \dot{f}_i^\alpha) \in \mathcal{T}\mathcal{F}_i^\alpha$ are the effort and flow variables and their time derivatives associated with input $\alpha$.
\end{definition}

As with other tangent-space quantities, the instantaneous power in tangent space is not associated with a physical power flow, but provides a mathematical construct for representing higher-order energy dynamics within a linear framework.
It is briefly introduced in~\cite{ilic2020unified} and further developed here.
\begin{definition}[Instantaneous power in tangent space]\label{Defn:PowerTangent}
For input $\alpha \in \{r,u,m\}$ (i.e., port, control, disturbance), the instantaneous power in tangent space into a component $\Sigma_i$ is,
\begin{align*}
    P_{t,i}^\alpha =  \frac{d e_i^\alpha}{dt} \frac{d f_i^\alpha}{dt} 
\end{align*}
where $\dot{e}_i^\alpha \in {T}_{e_i^\alpha}\mathcal{E}_i^\alpha$ and $\dot{f}_i^\alpha \in {T}_{e_i^\alpha}\mathcal{F}_i^\alpha$ respectively represent the time derivatives of the effort and flow variables associated with input $\alpha$.
\end{definition}

The effort and flow variables depend on the state and corresponding input,
\begin{align}\label{eqn:ef_pair}
e_i^\alpha = \varepsilon_i^\alpha(x_i, \alpha_i), \quad f_i^\alpha = \zeta_i^\alpha(x_i, \alpha_i), \quad \alpha \in \{r,u,m\},
\end{align}
where $\varepsilon_i^\alpha$ and $\zeta_i^\alpha$ are smooth maps. 
Typically, one of the variables is determined internally by the component dynamics, while the other is imposed via the interconnection.

\subsection{Energy space dynamic model}\label{sec:EnergyDefn}

Consider the energy state dynamics of a component $\Sigma_i$ introduced in~\eqref{eq:en_state_model}.
% , with energy state variables given by the stored energy $E_i$, its time derivative $p_i = \dot{E}_i$, and the stored energy in tangent space $E_{t,i}$. 
Throughout this formulation, we assume that the physical parameters of the component (e.g., resistance, inductance, capacitance) are constants.
\begin{subequations}\label{eq:en_state_model}
\begin{align} 
    \dot{E}_i &= - \frac{E_i}{\tau_i} + P_i^r + P_i^u + P_i^m, \label{eq:esp1} \\
    \dot{p}_i = \ddot{E}_i &= 4 E_{t,i} + 2\dot{Q}_{C,i} - \dot{Q}_i^r - \dot{Q}_i^u - \dot{Q}_i^m, \label{eq:esp2} \\
    \dot{E}_{t,i} &= -\frac{E_{t,i}}{\tau_{t,i}} + P_{t,i}^r + P_{t,i}^u + P_{t,i}^m, \label{eq:esp3}
\end{align}
\end{subequations}
where $E_i$, $p_i$, and $E_{t,i}$ denote the stored energy, its time derivative, and the tangent-space energy, respectively; $P_i^{(\cdot)}$, $\dot{Q}_i^{(\cdot)}$ and $P_{t,i}^{(\cdot)}$ represent the corresponding instantaneous power, reactive power rate and tangent-space power inputs; and $\dot{Q}_{C,i}$ denotes the rate of reactive power injected into the capacitor.
 
Equation~\eqref{eq:esp1} follows from energy conservation, balancing stored energy, dissipation, and external power inputs. 
Equation~\eqref{eq:esp2} relates the second derivative of stored energy to tangent space energy and reactive power rates; a derivation is provided in~\cite{ilic2018fundamental}.
Equation~\eqref{eq:esp3} describes tangent space energy dynamics, extending the formulation introduced in~\cite{ilic2020unified}.
Its structure mirrors~\eqref{eq:esp1} by applying conservation principles to higher-order quantities.

Building on~\cite{ilic2020unified}, we develop a third-order energy space model that models $E_{t,i}$ as a state variable.
Earlier second-order models~\cite{ilic2018multi,ilic2018fundamental,rupa2021feasible} treated $E_{t,i}$ as an exogenous term, which simplified analysis but was inconsistent with its physical role in the system’s energy evolution.
% By explicitly incorporating $E_{t,i}$ as a state, we introduce a third energy space input, tangent space power $P_{t,i}$, enhancing control flexibility, especially in systems with multiple inputs.

We now define \textit{interaction variables} that characterize energy exchanges. 
\begin{definition}[Interaction variables] The interaction variable for component $\Sigma_i$, denoted $z_i^\alpha$, for input $\alpha \in \{r,u,m\}$, is defined as,
\begin{align*}
    z^\alpha_i = \begin{bmatrix} 
        \int_0^t P^\alpha_i(\tau)\, d\tau \\
        \int_0^t \dot{Q}^\alpha_i(\tau)\, d\tau \\
        \int_0^t P^\alpha_{t,i}(\tau)\, d\tau
    \end{bmatrix}.
\end{align*}
This variable remains constant, i.e., $\dot{z}^\alpha_i = 0$,
when the corresponding interconnection with the component is removed.
\end{definition}

The \textit{port interaction variable} $z_i^r$ captures the accumulated energy interactions at the port and is therefore a natural candidate for the common output that is communicated for distributed control. 
Its existence and structural properties are formally established in~\cite{Ilic1993Linear}.
Its elements obey generalized Tellegen’s theorem~\cite{tellegen} and satisfy the interconnection feasibility condition stated in Problem~\Pref{energy_control_objectives}.

The \textit{control interaction variable} $z_i^u$ and the \textit{exogenous interaction variable} $z_i^m$ characterize the effects of control and disturbance inputs, respectively. 
These do not serve as common outputs and are not communicated to neighboring components; they are solely used for analytical convenience.

\section{Distributed control problem formulation in energy space}
\label{sec:distributed_energy}
In this section, we use the energy space modeling framework introduced in Section~\ref{sec:energy_modeling} to formulate a distributed component-level control problem in energy space that satisfies the original control objectives defined in the conventional state space. 
Control inputs are designed in energy space using local energy space states and communicated interaction rates, and then mapped to a physically implementable control in the conventional space. 
To achieve this, we introduce a multilayered control architecture.

\subsection{Multilayered distributed control architecture} 
\label{subsec:layer}
We introduce a two-layer modeling structure for each component $\Sigma_i$ in the interconnected system that maintains consistency between control synthesis in energy space and implementation in the physical system.
The lower layer, or \textit{physical space model}, represents the system using conventional state variables, as described in~\eqref{Eqn:GeneralCompModel_COnv}.
It captures the nonlinear dynamics as they evolve under control, disturbances, and port inputs.
The upper layer, termed the \textit{energy space model} and introduced in Section~\ref{sec:EnergyDefn}, maps the component dynamics into a linear form in terms of energy state and interaction variables. 
This linear structure enables systematic control synthesis, performance guarantees, and a physically intuitive  framework for modeling and control.
The resulting energy space control laws are subsequently mapped back to the lower physical layer for implementation. 

The energy state variable for $\Sigma_i$ is defined as $x_{z,i} = \begin{bmatrix} E_i & p_i & E_{t,i} \end{bmatrix}^\top$
and the energy space dynamics are given by,
\begin{subequations}
\label{Eqn:GeneralCompModel_Energy}
\begin{align}
&\text{Initial condition:} \quad x_{z,i}(0) = x_{z,i,0}, \\
&\text{Energy dynamics:} \quad \dot{x}_{z,i} = A_{z,i} x_{z,i} + B_{t,i} \Qd_{C,i} + B_{z,i}(\zd_i^r + \zd_i^u + \zd_i^m), \label{eq:energy_dynamics} \\
&\text{Energy output:} \quad \mathbf{y}_{z,i} = \begin{bmatrix} E_i \\ p_i \end{bmatrix}, \label{eqn:ydef}\\
&\text{Common output:} \quad
 \zd_i^r = \phi_{z,i}(x_{z,i}, \Qd_{C,i}, z_i^u, z_i^m), \quad z_i^r(0) = z_{i,0}^r. \label{eqn:zdr_func}
\end{align}
\end{subequations}
The energy state dynamics evolve under the system matrix $A_{z,i}$, with inputs from the rate of change of port, control, and disturbance interaction variables ($\dot{z}_i^r$, $\dot{z}_i^u$, $\dot{z}_i^m$), and an internal variable $\Qd_{C,i}$ capturing rate of reactive power in a capacitor.
The system matrix $A_{z,i}$, along with the input matrices $B_{z,i}$ and $B_{t,i}$ are given by,
\begin{align*}
    A_{z,i} = \begin{bmatrix}
        -\frac{1}{\tau_i} & 0 & 0\\ 
        0 & 0 & 4\\
        0 & 0 & -\frac{1}{\tau_{t,i}}
    \end{bmatrix}, 
    \quad B_{t,i} = 2, 
    \quad B_{z,i} = 1.
\end{align*}
$\mathbf{y}_{z,i}$ is the local energy space output of interest for which the energy space control is designed. 
The rate of change of port interaction variable $\zd_i^r$ is chosen as the common output communicated to neighboring components for distributed control, as it satisfies the interconnection feasibility condition implicitly by the generalized Tellegen’s theorem~\cite{tellegen}.

We now state the assumptions used in this work. Assumption~\ref{ass:equilibrium} ensures the existence of at least one state trajectory that satisfies the output tracking objective under the given inputs.
\begin{assumption} \label{ass:equilibrium}
Given admissible port, control, and exogenous input trajectories
$r_i(\cdot)$, $u_i(\cdot)$, and $m_i(\cdot)$, there exists a non-empty set of state trajectories $\mathcal{X}_{\text{eq},i}(t) \subseteq \mathcal{X}_i$ such that, for all $t \ge 0$,
\begin{align*}
    x_i(t) \in \mathcal{X}_{{eq},i}(t)
    \quad \text{and} \quad
    f_{y,i}\!\left(x_i(t), u_i(t), r_i(t), m_i(t)\right)
    = y_i^{{ref}}(t).
\end{align*}
\end{assumption}
In this work, we restrict attention to \textit{components} with a single output and a single control input. 
To accommodate this single degree of freedom within the energy space formulation, we introduce a scalar \emph{primary control input} $u_{z,i}$ in the upper layer.
Assumption~\ref{ass:control_interaction_map} specifies how the control interactions depend on $u_{z,i}$ through a technology-dependent mapping.
\begin{assumption}
\label{ass:control_interaction_map}
The rate of change of the control interaction is given by a smooth map
$U_i : \mathbb{R} \times \mathcal{X}_i \to \mathbb{R}^3$,
\begin{align}\label{eqn:zdu_uz}
    \dot z_i^u = U_i(u_{z,i}, x_i),
\end{align}
where $u_{z,i} \in \mathbb{R}$ is the primary control input in energy space and
$x_i \in \mathcal{X}_i$ is the component’s physical state.
\end{assumption}
Assumption~\ref{ass:uz_to_ui_map} ensures that control inputs synthesized in energy space admit a physically implementable realization in the physical system.
\begin{assumption}
\label{ass:uz_to_ui_map}
There exists a smooth, uniquely defined mapping \\$d_i:\mathbb{R}\times\mathcal{X}_i\to\mathcal{U}_i$ that maps the energy space primary control to the physical control as,
\begin{align}\label{eqn:pc_to_ec}
u_i = d_i^u(u_{z,i}, x_i).
\end{align}
\end{assumption}
Although the energy-space output $\mathbf{y}_{z,i}$ is two-dimensional, it encodes the same single degree of freedom as the physical output $y_i$ through the constraint $p_i = \dot{E}_i$. 
Assumption~\ref{ass:upper_layer_output_lift} enforces that a smooth mapping from the physical output reference to the corresponding energy-space output reference exists.
\begin{assumption}
\label{ass:upper_layer_output_lift}
For each component $\Sigma_i$, there exists a smooth map $\Phi_{\dot{\mathcal{Z}}_i^r}$ s.t., 
\begin{align}\label{eq:output_reference_map}
    \mathbf{y}_{z,i}^{{ref}}(t) = \Phi_{\dot{\mathcal{Z}}_i^r}\!\left(y_i^{{ref}}(t)\right), 
    \quad \dot{\mathcal{Z}}_i^r := \{\dot z_j^r \mid j \in \mathcal{C}_i\},
\end{align} 
where the map may depend on $\zd_j^r$ communicated by neighboring components. 
\end{assumption}

\begin{figure*}
    \centering
    \includegraphics[width=\linewidth]{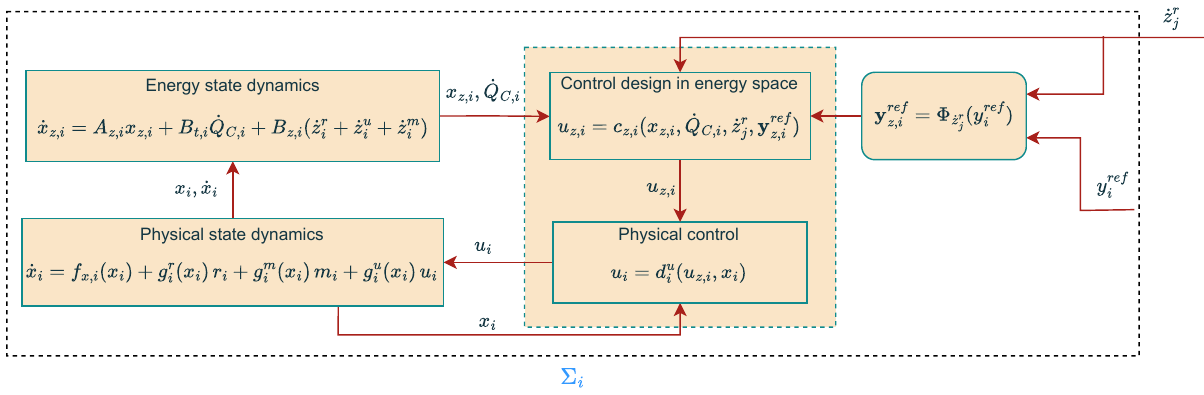} 
    \caption{Multilayered distributed control architecture for a component $\Sigma_i$}
    \label{fig:multilayered_control_architecture}
\end{figure*}

For each component $\Sigma_i$, the multilayered architecture operates by measuring the physical state, lifting it to energy space, designing an energy space control input using local and neighbor information, and mapping it back to the physical space for implementation, as shown in Figure~\ref{fig:multilayered_control_architecture}. 
\begin{enumerate}
    \item \textbf{State measurement:} 
    $\Sigma_i$ measures its physical state $x_i$.
    \item \textbf{Lifting to energy space:} 
    $\Sigma_i$ computes its energy state $
        x_{z,i}$ and internal variable $\Qd_{C,i}$, if a capacitor is present.
%     \item \textbf{Energy space control design:} 
%     \newline Each component $\Sigma_i$ designs its primary control input $u_{z,i}$ based on available system information, including:
% \begin{itemize}
%     \item the local energy space state $x_{z,i}$ and internal variable $\Qd_{C,i}$,
%     \item the port interaction variables $z_j^r$ received from neighboring components $\Sigma_j$, where by the feasibility condition in Problem~\Pref{prob:control_objectives}, the local rate of change $\zd_i^r$ can be computed as $\zd_i^r = -\sum_{j \in \mathcal{C}_i} \zd_j^r$,
%     \item and the energy space output reference $\mathbf{y}_{z,i}^{{ref}}$, computed via~\eqref{eq:output_reference_map}.
% \end{itemize}
% Using this information, the higher level primary control input $u_{z,i}$ is determined as a smooth function of the local energy state, internal variable, interaction information, and reference trajectory,
% \begin{align}
%     u_{z,i} = c_{z,i}(x_{z,i}, \Qd_{C,i}, \zd_i^r, \mathbf{y}_{z,i}^{{ref}}), \label{eqn:cui}
% \end{align}
% where $c_{z,i}(\cdot)$ denotes the control law that maps the available system information to the energy space control input $u_{z,i}$.
\item \textbf{Energy space control design:}
The primary control input $u_{z,i}$ is computed using available information such as, energy state $x_{z,i}$ and internal variable $\Qd_{C,i}$, rate of port interactions ${\zd}_j^r$ received from neighboring components, and energy space output reference $\mathbf{y}_{z,i}^{{ref}}$,
\begin{align*}
    u_{z,i} = c_{z,i}(x_{z,i}, \Qd_{C,i}, \dot{\mathcal{Z}}_i^r, \mathbf{y}_{z,i}^{{ref}}), \quad \dot{\mathcal{Z}}_i^r := \{\dot z_j^r \mid j \in \mathcal{C}_i\} .
\end{align*}
% where $c_{z,i}(\cdot)$ denotes the control law that maps the available system information to the energy space control input $u_{z,i}$.
    \item \textbf{Mapping to physical control input:} 
    The primary control input $u_{z,i}$ is mapped to the physical control input $u_i$ via~\eqref{eqn:pc_to_ec} and implemented.
    % If needed, $u_{z,i}$ can be used to compute the control interaction variables via the mapping in~\eqref{eqn:zdu_uz}.
\end{enumerate}
At the next time step, after updating $x_i$, 
component $\Sigma_i$ computes $\zd_i^r$ using~\eqref{eqn:zdr_func} or an observer, and communicates it to neighboring components.

\subsection{Distributed control problem posing in energy space} 
\label{subsec:energy_control_objectives}

Building on the general distributed control framework in Problem~\ref{prob:control_objectives}, we now pose the corresponding control problem in energy space.
\begin{problem}[Distributed control problem in energy space]\label{energy_control_objectives}
Given an interconnected system consisting of a set $\mathcal{N}$ of components defined in energy space~\eqref{Eqn:GeneralCompModel_Energy},
the objective is to design a smooth distributed feedback control law,
\begin{align}\label{eqn:cui}
    u_{z,i} = c_{z,i}(x_{z,i}, \mathbf{y}^{{ref}}_{z,i}, \Qd_{C,i}, \{{\zd}_j^{r}\}_{j \in \mathcal{C}_i}), \;\; \forall i \in \mathcal{N},
\end{align}
which depends on the local energy state $x_{z,i}$, internal variable $\Qd_{C,i}$, minimal information $\zd_j^r$ communicated by neighboring components $\Sigma_j$, and the energy space output reference $\mathbf{y}^{{ref}}_{z,i}$, such that the following is satisfied:
\begin{enumerate}
    \item {Interconnection constraint:}
    The common output variables of neighboring components must satisfy,
    \begin{align} \label{eqn:sumj_i}
        \zd_i^r = -\sum_{j \in \mathcal{C}_i} \zd_j^r, \quad \forall i \in \mathcal{N} .
    \end{align}
    (This condition is satisfied by the choice of port interaction rates as common output due to generalized Tellegen’s theorem~\cite{tellegen}.)
    \item {Control objective:}
    The energy space output $\mathbf{y}_{z,i}$ must track a consistent reference $\mathbf{y}^{{ref}}_{z,i}$ given by the smooth map $\Phi_{\dot{\mathcal{Z}}_i^r}$ as defined in~\eqref{eq:output_reference_map},
\begin{align*}
    \mathbf{y}_{z,i}(t) \to \mathbf{y}^{{ref}}_{z,i}(t) \quad \text{as } t \to \infty, \quad \forall i \in \mathcal{N}.
\end{align*}
\end{enumerate}
\end{problem}

\begin{theorem}[Inter-layer consistency for output tracking]
\label{thm:interlayer_consistency}

Suppose that, for each component $\Sigma_i$:
\begin{enumerate}
    \item The energy space and physical outputs satisfy,
    \begin{align*}
        \mathbf{y}_{z,i} = \mathbf{y}_{z,i}^{{ref}}
    \;\Longrightarrow\;
    y_i = y_i^{{ref}}.
    \label{eq:yz_to_y}
    \end{align*}
    \item The common output is chosen as the port interaction rates~\eqref{eqn:zdr_func}.
\end{enumerate}
Then, solving Problem~\Pref{energy_control_objectives} solves Problem~\Pref{prob:control_objectives}.
\end{theorem}

\begin{remark}
Condition (1) in Theorem~\ref{thm:interlayer_consistency} is component-dependent. In many practical systems, the physics ensure that energy space output tracking implies physical output tracking; illustrative examples are provided in Section~\ref{sec:examples}.
\end{remark}

% \begin{remark}\label{remark:system_level_coord}
% While $\dot z_i^r$ can be computed locally, two-way communication with neighboring components is essential for system-level coordination. 
% The design of system-level control laws that allocate port interaction references $\dot z_i^{r,{ref}}$ is beyond the scope of this work. 
% In this paper, we assume that the aggregate rate of port interactions received from neighboring components satisfies
% \begin{align*}
%     \sum_{j \in \mathcal{C}_i} \dot z_j^r = \dot z_i^{r,{ref}}.
% \end{align*}
% Under this assumption, we focus on establishing local stability properties of each component given the resulting interactions.
% \end{remark}

\section{Energy-based component-level control}
\label{sec:energy-based_component-level_control}

To solve Problem~\Pref{energy_control_objectives}, we design a generalized higher-level control using established nonlinear control techniques~\cite{khalil2002nonlinear}, namely feedback linearization control (FBLC) and sliding mode control (SMC).
The primary control input is selected as $u_{z,i} := \dot Q_i^u$, with other control interactions determined via~\eqref{eqn:zdu_uz}.
As defined in~\eqref{eqn:cui}, $u_{z,i}$ is a function of the local energy state, internal variable, communicated interactions, and the energy space output reference.
Given communicated neighbor interactions $\zd_j^r$, each component computes its local interaction variable $\zd_i^r$ using~\eqref{eqn:sumj_i}.
Any estimation errors in the communicated port interaction variables are absorbed into $\zd_i^m$.

Using~\eqref{eq:esp1}–\eqref{eq:esp2}, the energy space tracking error dynamics become,
\begin{align}
    \dot{\mathbf{y}}_{z,i} - \dot{\mathbf{y}}_{z,i}^{ref} & = \begin{bmatrix}
        p_i - p_i^{ref} \\ 4 \Eti + 2\Qd_{C,i} - \Qdi{r} - \Qdi{u} - \Qdi{m} - \pd_i^{ref} \end{bmatrix}. \label{eq:iodyn}
\end{align}

\subsection{Energy-based feedback linearization control}
% Feedback linearization control (FBLC) is a nonlinear control technique~\cite{khalil2002nonlinear} that cancels system nonlinearities by applying an exact transformation through the control input. 
To ensure $\mathbf{y}_{z,i}$ tracks $\mathbf{y}^{ref}_{z,i}$, we propose the following energy-based FBLC,
\begin{align}
    u_{z,i}^{fblc} &= 4 \Eti + 2\dot{Q}_{C,i} + \sum_{j \in \mathcal{C}_i}\Qd_{j}^r + K_1 (E_i -E_{i}^{ref}) + K_2 (p_i - p_i^{ref}) - \pd_i^{ref} \label{eqn:FBLC}
\end{align}
where $K_1, K_2 > 0$ are constant gains.
Clearly, $u_{z,i}^{fblc}$ depends on the available local and communicated information, as defined in~\eqref{eqn:cui}. 
However, we note that $u_{z,i}^{{fblc}}$ also depends on the term $\dot{p}_i^{{ref}}$, which may not be directly available and can be estimated in practice.  

Substituting \eqref{eqn:FBLC} into \eqref{eq:iodyn} and applying~\eqref{eqn:sumj_i} yields,
\begin{align}
    \dot{\mathbf{y}}_{z,i} - \dot{\mathbf{y}}_{z,i}^{ref} &= \begin{bmatrix}
        y_{z,i}^2 - y_{z,i}^{2, ref}\\ - K_1 (y_{z,i}^1 - y_{z,i}^{1,ref}) - K_2 (y_{z,i}^2  - y_{z,i}^{2,ref}) - \Qdi{m}
    \end{bmatrix}, \label{eqn:pdclosed_FBLC}
\end{align}
where $y_{z,i}^1 = E_i$ and $y_{z,i}^2 = p_i$.
% where, $\tilde{\Qd}_i := \Qdi{in} - \Qdi{out}$ is the difference in the incoming and outgoing $\Qdi{}$ as given in \eqref{eqn:delay_Qtilde}.
We now establish sufficient conditions for asymptotic convergence of the energy space output tracking error.

\begin{theorem}[Performance with energy-based FBLC]\label{Theorem:CtrlFBLC}
Consider the virtual control $u_{z,i} = \dot{Q}^u_i$ in~\eqref{eqn:FBLC}.
The closed-loop energy space output dynamics \eqref{eqn:pdclosed_FBLC} are asymptotically stable if,
\begin{align}
    |\Qdi{m}| \leq  K_2 |p_i - p_i^{ref}|. \label{eq:ass_FBLC}
\end{align}
% where $\Tilde{\dot{Q}}_i := \Qd^{in}_i - \Qd^{out}_i$. 
\begin{proof}
    Consider the candidate Lyapunov function,
\begin{align*}
        V_i^{fblc}(\mathbf{y}_{z,i} - \mathbf{y}_{z,i}^{ref}) &= \frac{1}{2} \left(K_1(y_{z,i}^1 - y_{z,i}^{1,ref})^2 \right. + \left.(y_{z,i}^2 - y_{z,i}^{2,ref})^2\right) .
\end{align*}
    % where $\mathbf{y}_{z,i}$ and $\mathbf{y}_{z,i}^{ref}$ are as defined in \eqref{eqn:ydef}.
    % $V_i^{fblc}$ is a valid Lyapunov function since $V_i^{fblc} = 0 $ if and only if $ \mathbf{y}_{z,i} - \mathbf{y}^{ref}_{z,i} = \mathbf{0}$, and $V_i^{fblc} > 0$ otherwise.
    % It is thus positive definite and radially unbounded with respect to the tracking error.
    Taking its time derivative and substituting~\eqref{eqn:pdclosed_FBLC} yields,
    \begin{align}
        \dot{V}_i^{fblc} &= K_1 \left(y_{z,i}^1 - y_{z,i}^{1,ref}\right)\left(y_{z,i}^2 - y_{z,i}^{2,ref}\right)  \nonumber\\
        & \hspace{1cm} + \left(y_{z,i}^2 - y_{z,i}^{2,ref}\right)\left(- K_1 (y_{z,i}^1 - y_{z,i}^{1,ref}) - K_2 (y_{z,i}^2 - y_{z,i}^{2,ref}) - \Qdi{m} \right), \nonumber\\
        &= -K_2\left(y_{z,i}^2 - y_{z,i}^{2,ref}\right)^2 - \Qdi{m}\left(y_{z,i}^2 - y_{z,i}^{2,ref}\right).  \label{eq:vd2}
    \end{align}
    Applying the condition~\eqref{eq:ass_FBLC}, we obtain $ \dot{V}_i^{fblc} \leq 0$.
    This proves Lyapunov stability; to prove asymptotic stability, we use LaSalle's invariance principle.
    Equation \eqref{eq:vd2} indicates that for arbitrary $\Qdi{m}$ values under~\eqref{eq:ass_FBLC}, the largest invariant set of points $\mathbf{y}_{z,i} \in \mathbb{R}$ that satisfies $\dot{V}_i^{fblc} = 0$ is the equilibrium $ \mathbf{y}_{z,i} = \mathbf{y}_{z,i}^{ref}$.
    This is because if $ y_{z,i}^1 \neq y_{z,i}^{1,ref}$ or $y_{z,i}^2 \neq y_{z,i}^{2, ref}$ then by equation \eqref{eqn:pdclosed_FBLC} $\dot{y}_{z,i}^2 \neq \dot{y}_{z,i}^{2,ref}$, violating invariance.
    Hence, the closed-loop energy space output dynamics are asymptotically stable.  
\end{proof}
\end{theorem}

Condition~\eqref{eq:ass_FBLC} implies that the tracking error in $p_i$ is bounded by $|\dot Q_i^m|/K_2$. 
Thus, reducing disturbances improves FBLC tracking performance.

\subsection{Energy-based sliding mode control (SMC)}
% Sliding mode control (SMC) is a robust nonlinear control technique~\cite{khalil2002nonlinear} that drives system trajectories onto a prescribed sliding surface in finite time and maintains them there, ensuring robustness to model uncertainties and disturbances. 
For the energy-based SMC design, consider the sliding surface, 
\begin{align*}
    \sigma_i := p_i - p_i^{ref} + M_1 (E_i -E_{i}^{ref}) = 0.
\end{align*}  
% On this surface, asymptotic convergence of the energy space output tracking error is ensured.
We propose the following energy-based SMC,
\begin{align}
    u_{z,i}^{smc} &= 
    4 \Eti + 2\dot{Q}_{C_{i}} + \sum_{j \in \mathcal{C}_i}\Qd_{j}^r + M_1 (p_i - p_i^{ref}) + M_\circ \sign(\sigma_i) - \pd_i^{ref}, \label{eqn:SMCcontrol}
\end{align}
where $M_\circ, M_1 > 0$ are gain constants.
Similar to the energy-based FBLC, $u_{z,i}^{smc}$ is constructed only from the available local and communicated information, as defined in~\eqref{eqn:cui}.

Substituting \eqref{eqn:SMCcontrol} into \eqref{eq:iodyn} and applying~\eqref{eqn:sumj_i} yields,
\begin{align}
    &\dot{\mathbf{y}}_{z,i} - \dot{\mathbf{y}}_{z,i}^{ref} = \begin{bmatrix}
        p_i - p_i^{ref} \\ 
     - M_\circ\sign(\sigma_i) - M_1 (p_i - p_i^{ref}) - \Qdi{m}
    \end{bmatrix} . \label{eqn:pdclosed_SMC}
\end{align}
We now establish sufficient conditions for finite-time convergence to the sliding surface and asymptotic convergence of the tracking error.

\begin{theorem}[Performance with energy-based SMC]\label{Theorem:CtrlSMC}
Consider the virtual control $u_{z,i} = \Qdi{u}$ in~\eqref{eqn:SMCcontrol}.
The closed-loop dynamics \eqref{eqn:pdclosed_SMC} reach the sliding surface $\sigma_i = 0$ in finite time,  
\begin{align}\label{eq:tr}
    t_r \leq \frac{|\sigma_i(0)|}{M_\circ - \bar{M}},
\end{align} provided that,
    \begin{align}
        |\Qdi{m}| \leq \Bar{M} < M_\circ, \label{eq:ass_SMC}
    \end{align}
    where $\Bar{M}:= \sup |\Qdi{m}(t)|$ and $\sigma_i := p_i - p_i^{ref} + M_1 (E_i -E_{i}^{ref})$.
    Moreover, tracking error converges asymptotically on the sliding surface.
\begin{proof}
    Consider the Lyapunov function, $V_i=\frac{1}{2}\sigma_i^2$.
    Using~\eqref{eqn:pdclosed_SMC},
    \begin{align*}
        \dot{V}_i^{smc} 
         &= \sigma_i \left(\pd_i - \pd_i^{ref} + M_1 (p_i - p_i^{ref})\right),\\
         % &= \sigma_i \left( - M_\circ \sign(\sigma_i) - \Qdi{m}\right),\\
         &= - M_\circ |\sigma_i| -  \Qdi{m} \sigma_i.
    \end{align*}
    Applying the condition~\eqref{eq:ass_SMC}, we obtain,
    \begin{align}
        \dot{V}_i^{smc} \leq - (M_\circ - \Bar{M}) |\sigma_i| < 0, \label{eqn:VSMC}
    \end{align}
    showing that $\dot{V}_i^{{smc}}$ is negative definite. 
    Integrating~\eqref{eqn:VSMC}, we find,
    \begin{align*}
        % \dot{V}_i^{smc} &\leq - (M_\circ - \Bar{M}) \sqrt{2 V_i^{smc}},\\
        \sqrt{2V_{i}^{smc}(t_r)} - \sqrt{2V_i^{smc}(0)} &\leq -(M_\circ - \Bar{M})t_r,\\
        |\sigma_i(t_r)| - |\sigma_i(0)| &\leq -(M_\circ - \Bar{M})t_r.
    \end{align*}
    Setting $\sigma_i(t_r) = 0$ gives~\eqref{eq:tr}.
    Once the system reaches the sliding surface,
    \begin{align*}
        p_i = -M_1 (E_i - E_{i}^{{ref}}) + p_i^{{ref}},
    \end{align*}
    ensuring asymptotic convergence of the energy space output tracking error.        
\end{proof}
\end{theorem}
Condition \eqref{eq:ass_SMC} is less conservative than~\eqref{eq:ass_FBLC}, as it only requires a constant bound on the disturbance $|\Qdi{m}|<M_\circ$ for convergence.
Unlike FBLC, SMC guarantees finite reaching time. 
However, high-frequency switching in SMC may induce chattering, requiring smoothing in practice.

\section{Examples}
\label{sec:examples}

This section evaluates the proposed energy-based distributed control methods through two examples.
The first considers voltage regulation in an inverter-driven RLC circuit supplying constant and time-varying loads. 
The second examines frequency regulation of a synchronous generator under a smooth step load variation.
In both cases, energy-based FBLC and SMC are compared with conventional benchmark controllers.

\subsection{Voltage regulation in an RLC circuit}
\begin{figure}[ht]
    \centering
    \includegraphics[width=0.6\linewidth]{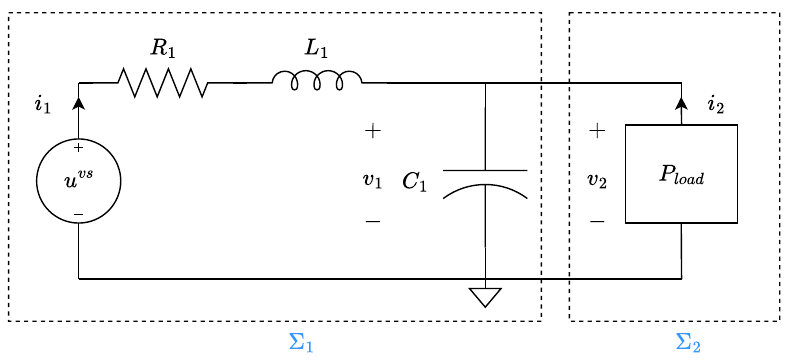}
    \caption{RLC circuit with an inverter-controlled voltage source $u^{vs}$ ($\Sigma_1$) supplying power to a black-box with unknown internal dynamics ($\Sigma_2$)}
    \label{fig:RLCcircuit}
\end{figure}
Figure~\ref{fig:RLCcircuit} shows an RLC circuit with an inverter-controlled voltage source $u^{vs}$ supplying power to a load with specified power demand $P_{load}$.
The objective is to regulate the terminal voltage $v_1$ to $80~\text{V}$.
% The system is modeled as two interacting subsystems: the source $\Sigma_1$ and the load $\Sigma_2$, consistent with the general structure in Figure~\ref{fig:system_general}.
The physical state, control and output variables of $\Sigma_1$ are,
\begin{align*}
    x_1 = \begin{bmatrix}
        i_1 \\
        v_1
    \end{bmatrix}, \quad u_1 = u^{vs}, \quad y_1 = v_1,
\end{align*}
where $i_1$ is the current flowing through the inductor $L_1$, and $v_1$ is the voltage across the capacitor $C_1$. 
$R_1$ represents the resistance in series with the inductor.
The lower level physical state space model for $\Sigma_1$ is given by,
\begin{align*}
    \frac{d}{dt}\underbrace{\begin{bmatrix}
        i_1 \\ v_1
    \end{bmatrix}}_{x_1} &= \underbrace{\begin{bmatrix}
        -\frac{R_1}{L_1} & -\frac{1}{L_1}\\
        \frac{1}{C_1} & 0 
    \end{bmatrix} \begin{bmatrix}
        i_1 \\ v_1
    \end{bmatrix}}_{f_{x,1}(x_1)} + \underbrace{\begin{bmatrix}
        \frac{1}{L_1} \\ 0
    \end{bmatrix}}_{g_1^u(x_1)} \underbrace{u^{vs}}_{u_1} + \underbrace{\begin{bmatrix}
        0 \\ -\frac{1}{C_1 v_1}
    \end{bmatrix}}_{g_1^r(x_1)} \underbrace{P_{load}}_{r_1} .
\end{align*}
To design the energy-based distributed controller for $\Sigma_1$, we select the relevant effort-flow pairs using~\eqref{eqn:ef_pair},
\begin{align*}
    &\left.
    \begin{aligned}
        e_1^u &= u_1, \quad &f_1^u &= i_1, \\
        e_1^C &= v_1, \quad &f_1^C &= C_1 \dot{v}_1,
    \end{aligned}
    \right\} &&\hspace{-1cm} \text{(for } \Sigma_1 \text{)} \\
    &\left.
    \begin{aligned}
        e_2^r &= v_2, \quad \;\; &f_2^r &= \frac{P_{load}}{v_2}.
    \end{aligned}
    \right. &&\hspace{-1cm} \text{(for } \Sigma_2 \text{)} 
\end{align*}
From the definitions in Section~\ref{sec:energy_modeling}, the load subsystem $\Sigma_2$ computes
$P_2^r$, $\dot{Q}_2^r$, and $P_{t,2}^r$, with $P_2^r = P_{load}$, and communicates it to $\Sigma_1$. 
Source subsystem $\Sigma_1$ computes its energy space vector, $x_{z,1} = \begin{bmatrix}
    E_1 & p_1 & E_{t,1}
\end{bmatrix}^\top$, and internal variable $Q_{C,1}$, using local state measurements.
% \begin{align*}
%     E_1 &= \frac{1}{2} L_1 i_1^2 + \frac{1}{2} C_1 v_1^2, \\
%     p_1 &= L_1 i_1 \frac{di_1}{dt} + C_1 v_1 \frac{dv_1}{dt}, \\
%     E_{t,1} &= \frac{1}{2} L_1 \left(\frac{di_1}{dt}\right)^2 + \frac{1}{2} C_1 \left(\frac{dv_1}{dt}\right)^2, \\
%  \Qd_{C,1} &= C_1 v_1 \frac{d^2 v_1}{dt^2} - C_1 \left( \frac{d^2 v_1}{dt^2} \right)^2.
% \end{align*}
%

The output of interest is the port voltage, which is to be regulated to a reference value of $y_1^{ref} = 80~\text{V}$. 
The corresponding higher level energy space output references are selected as,
\begin{align*}
    \mathbf{y}_{z,1}^{ref} &= \begin{bmatrix}
        E_1^{ref} \\ p_1^{ref}
    \end{bmatrix} = \begin{bmatrix}
        \frac{1}{2}L_1\left( \frac{P_{load}}{y_1^{ref}}\right)^2 + \frac{1}{2}C_1(y_1^{ref})^2
        \\L_1 \frac{P_{load}}{{y_1^{ref}}^2} \dot{P}_{load}
    \end{bmatrix}.
\end{align*}
If $\mathbf{y}_{z,1} = \mathbf{y}_{z,1}^{ref}$ holds for all admissible $P_{load}(t)$, 
then necessarily $y_1 = y_1^{ref}$, satisfying the inter-layer consistency conditions in Theorem~\ref{thm:interlayer_consistency}.

The primary control input $u_{z,1}$ is mapped to the physical control input $u_1$ through, 
\begin{align*}
    \frac{d u_1}{d t} = \frac{u_1}{i_1} \frac{d i_1}{d t} - \frac{u_{z,1}}{i_1}, \quad u_1(0) = 79~\text{V}.
\end{align*}
For all simulations and analysis, we use the following parameter values, 
\begin{align}
    R_1 = 100~\text{m}\Omega, \quad L_1 = 1.12~\text{mH}, \quad C_1 = 6.8~\text{mF}, \nonumber
\end{align}
with initial conditions, $i_1(0) = 12.8~\text{A}$ and $v_1(0) = 79~\text{V}$.

\subsubsection{Constant power load}
Consider a constant power load $P_{load} = 1~\text{kW}$.  
First, we consider a conventional proportional controller as our benchmark,
\begin{align}\label{eqn:prop_control}
    u^{vs}_p = u_{ref}^{vs} - K_{i}\left(i_1 - \frac{P_{load}}{y_1^{ref}}\right) - K_{v}(v_1 - y_1^{ref}),
\end{align}
where $u_{ref}^{vs}$ is the reference input applied at $\dot{v}_1 = 0$, 
\begin{align*}
    u_{ref}^{vs} = y^{ref}_1 + R_1\frac{P_{load}}{y_1^{ref}}. 
\end{align*}
The proportional control gains are chosen as $K_i = 5$ and $ K_v = 0.5$.
\begin{figure}[h]
    \centering
    % Subfigure 1: Current through inductor
    \begin{subfigure}[t]{0.48\textwidth}
        \centering
        \includegraphics[width=\textwidth]{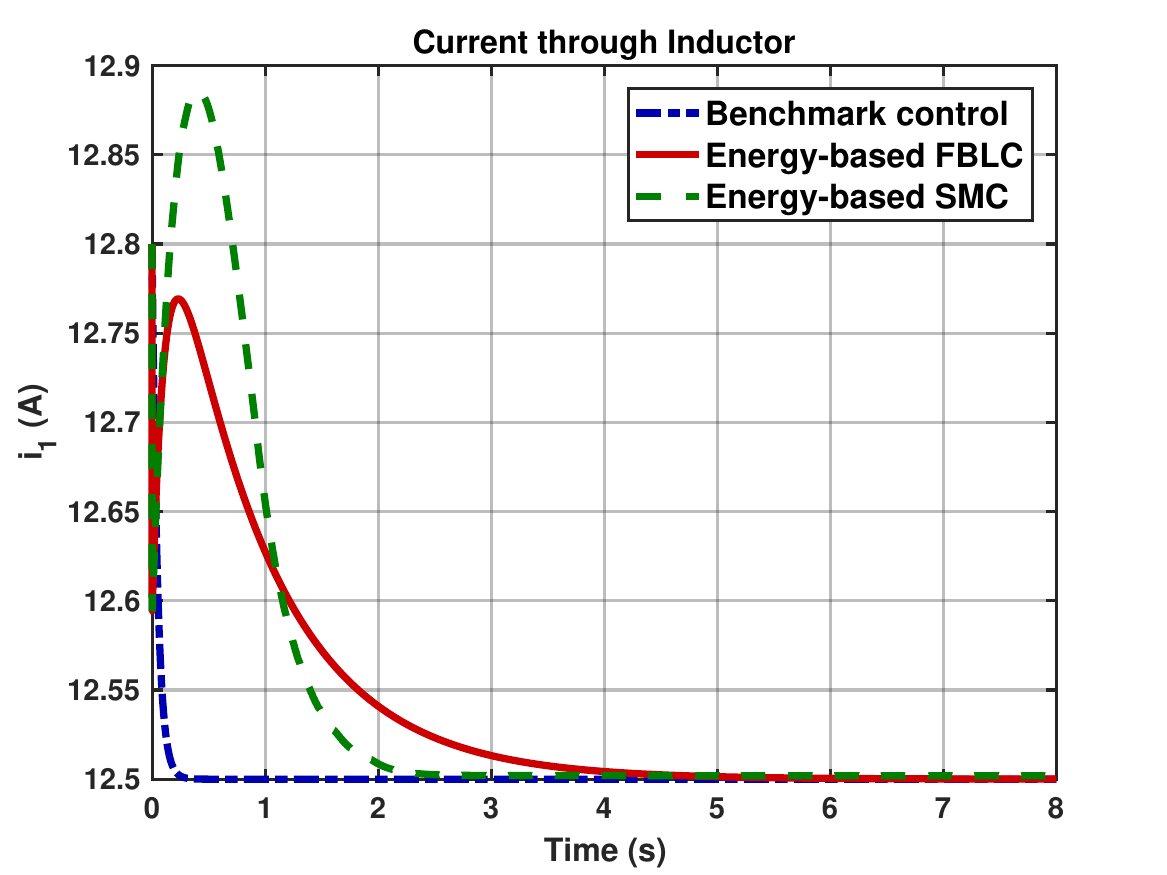}
        \caption{Current through inductor}
        \label{fig:I1}
    \end{subfigure}
    \hfill
    % Subfigure 2: Voltage across capacitor
    \begin{subfigure}[t]{0.48\textwidth}
        \centering
        \includegraphics[width=\textwidth]{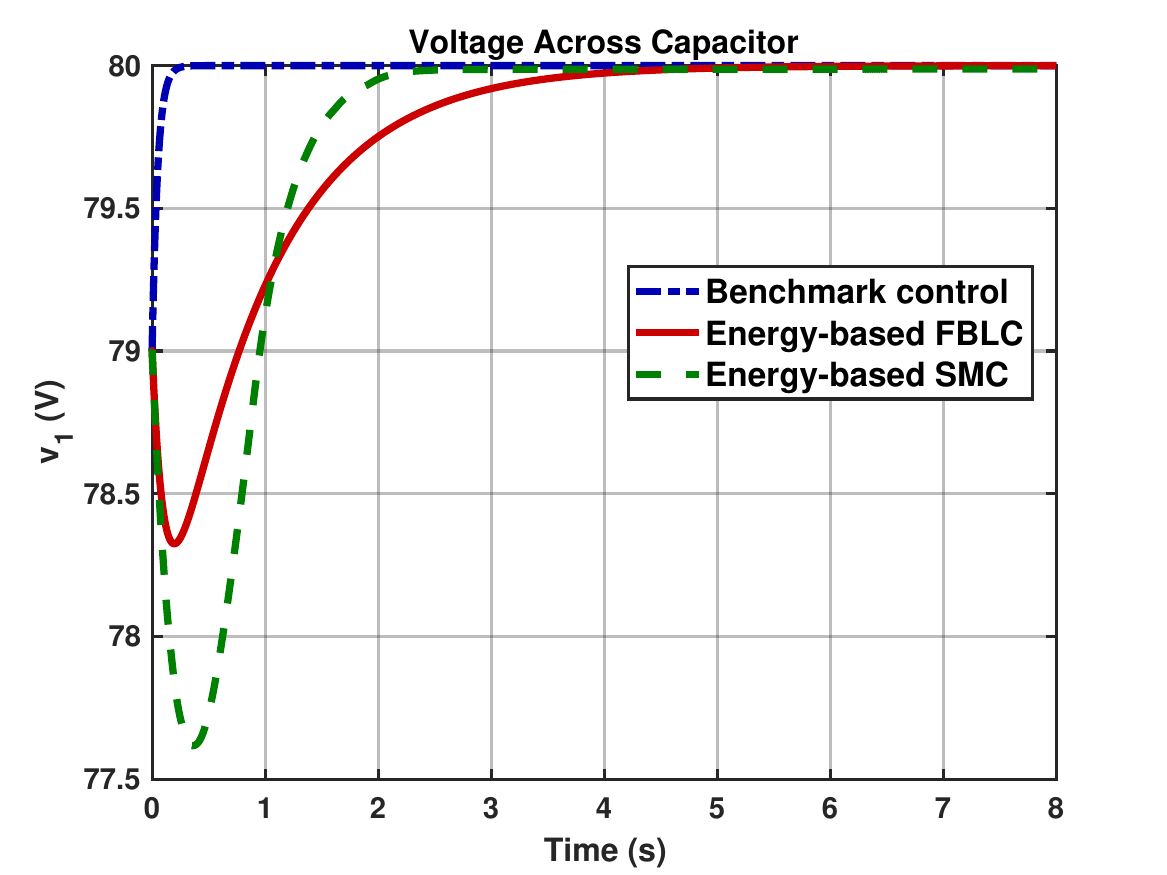}
        \caption{Voltage across capacitor}
        \label{fig:V1}
    \end{subfigure}
    \vspace{0.4em}
    % Subfigure 3: Applied control voltage
    \begin{subfigure}[t]{0.48\textwidth}
        \centering
        \includegraphics[width=\textwidth]{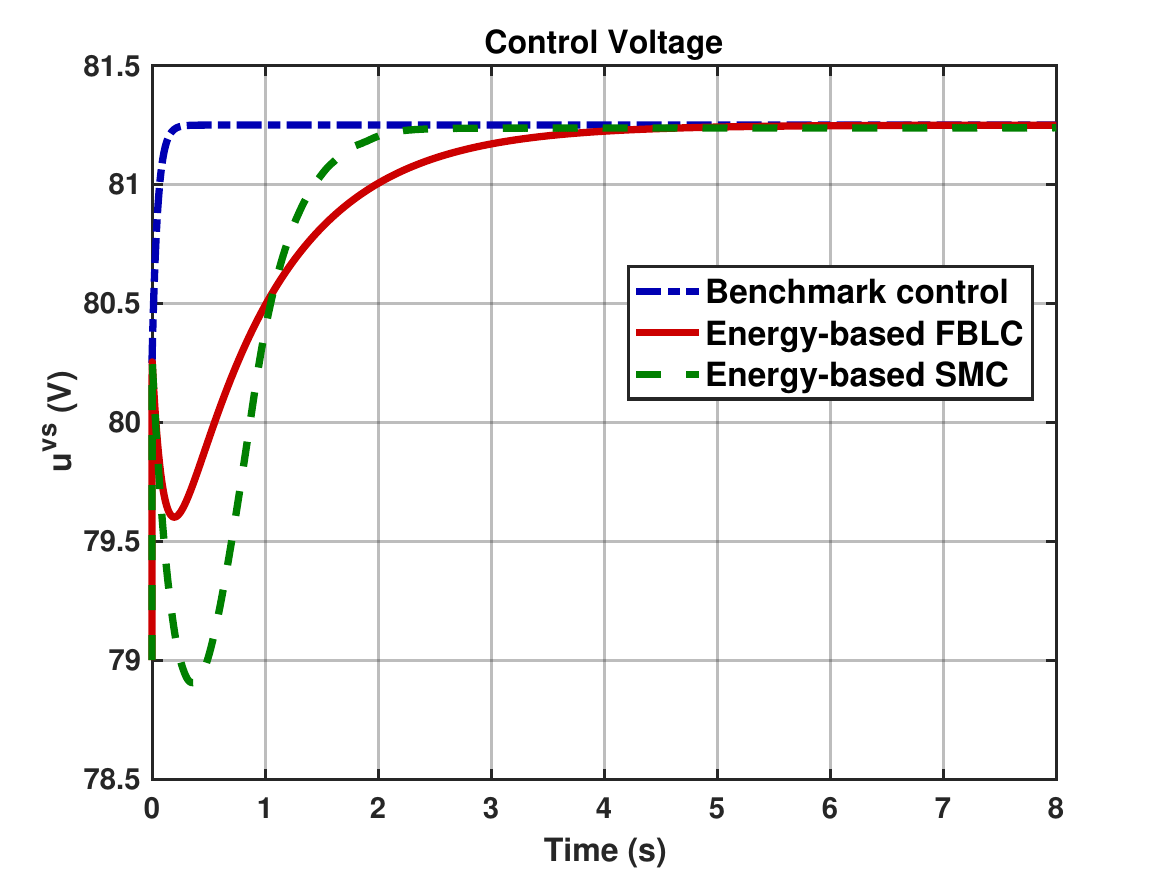}
        \caption{Applied control voltage}
        \label{fig:U}
    \end{subfigure}
    \caption{System response of the RLC circuit under proportional benchmark control \eqref{eqn:prop_control}, energy-based FBLC, and energy-based SMC for a constant power load of $1~\text{kW}$, with the objective of stabilizing and regulating the terminal voltage $v_1$ to $80~\text{V}$}
    \label{fig:IVUplot}
\end{figure}
 
Assuming no exogenous disturbances,
the energy-based FBLC gains are set as $K_1 = K_2 = 10$,
and the energy-based SMC gains as $M_\circ = 5.4$ and $M_1 = 2.9$.
Figure~\ref{fig:IVUplot} compares proportional control (in blue) with energy-based FBLC (in red) and SMC (in green). 
All methods regulate the terminal voltage to $80~\text{V}$, although proportional control relies on empirically tuned gains for stability. 
The FBLC yields smooth exponential convergence, while SMC achieves faster settling due to its finite-time reaching property.

\subsubsection{Time-varying power load}

\begin{figure}[h]
    \centering
    % Subfigure 0: TV load profile
    \begin{subfigure}[t]{0.48\textwidth}
        \centering
        \includegraphics[width=\textwidth]{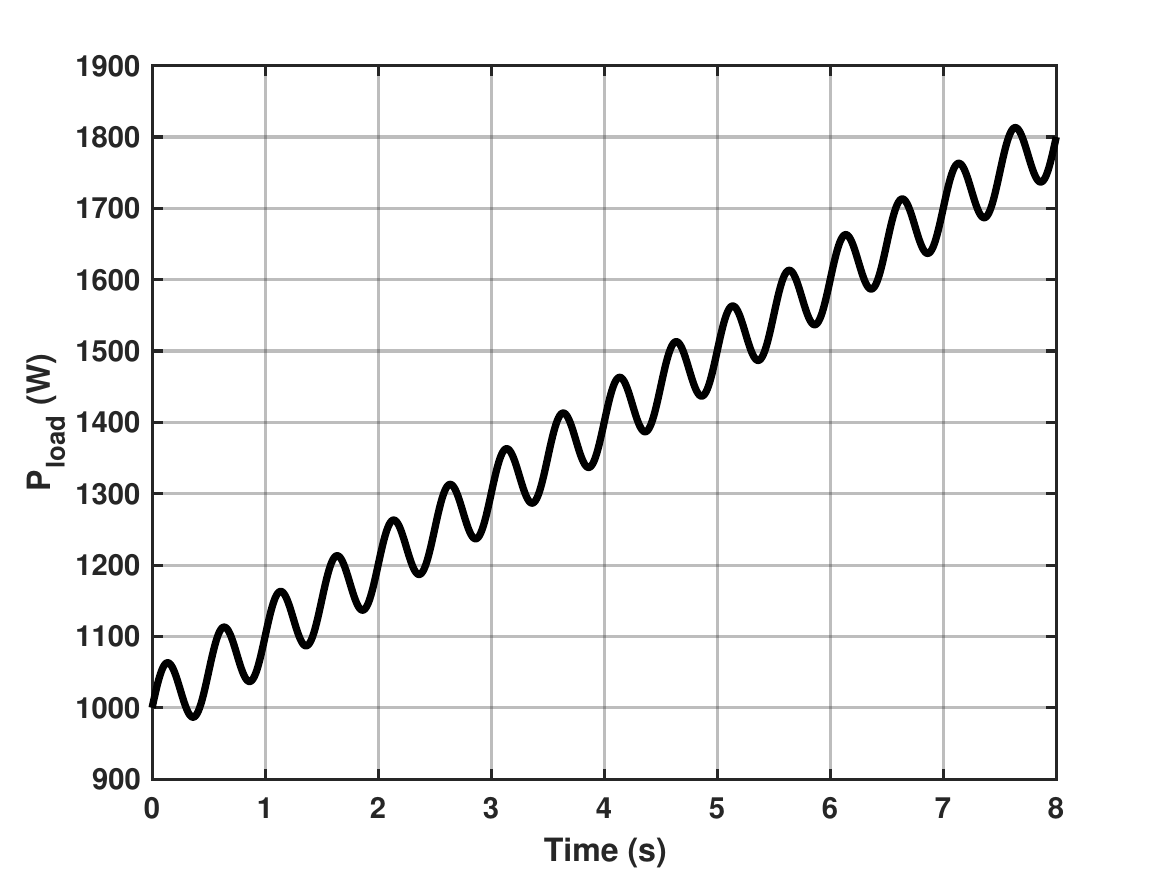}
        \caption{Time-varying load profile}
        \label{fig:load_profile_tv}
    \end{subfigure}
    \hfill
    % Subfigure 1: Current through inductor
    \begin{subfigure}[t]{0.48\textwidth}
        \centering
        \includegraphics[width=\textwidth]{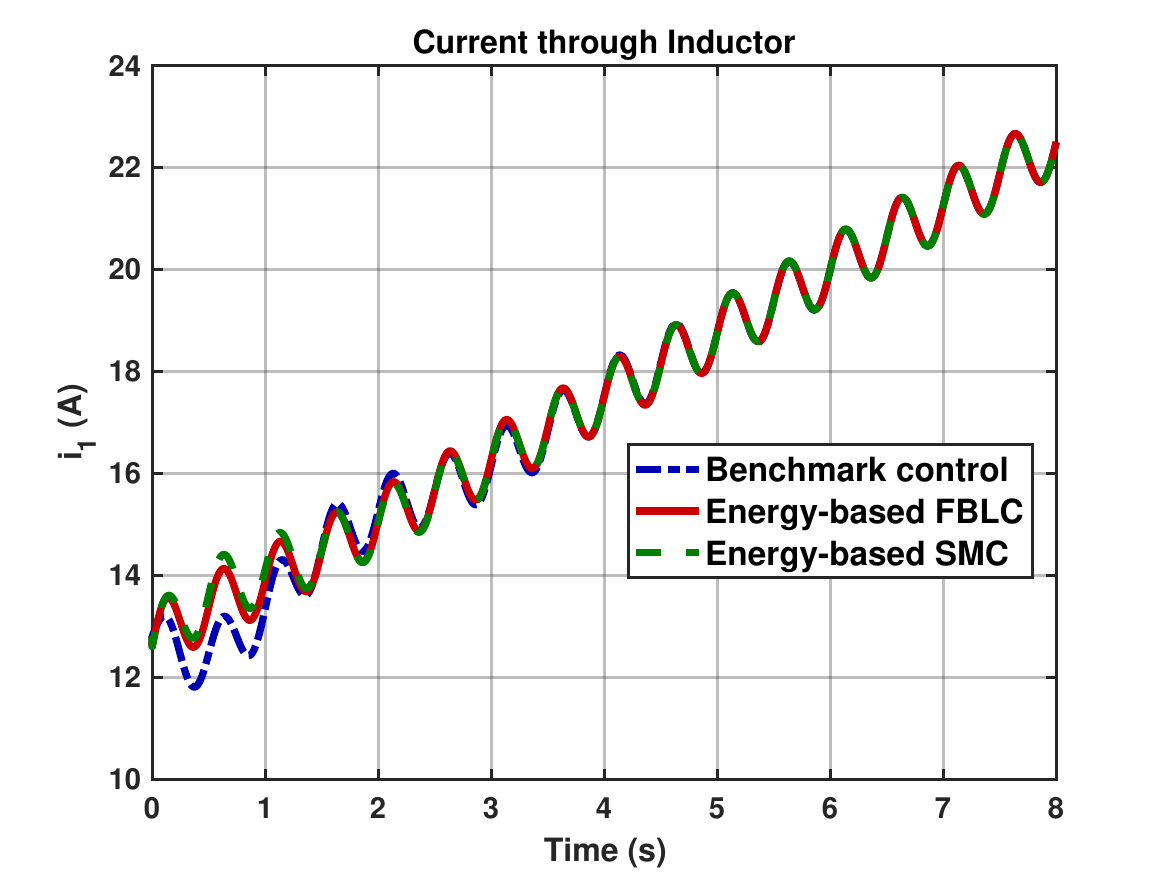}
        \caption{Current through inductor}
        \label{fig:I1_tv}
    \end{subfigure}
    \vspace{0.4em}
    % Subfigure 2: Voltage across capacitor
    \begin{subfigure}[t]{0.48\textwidth}
        \centering
        \includegraphics[width=\textwidth]{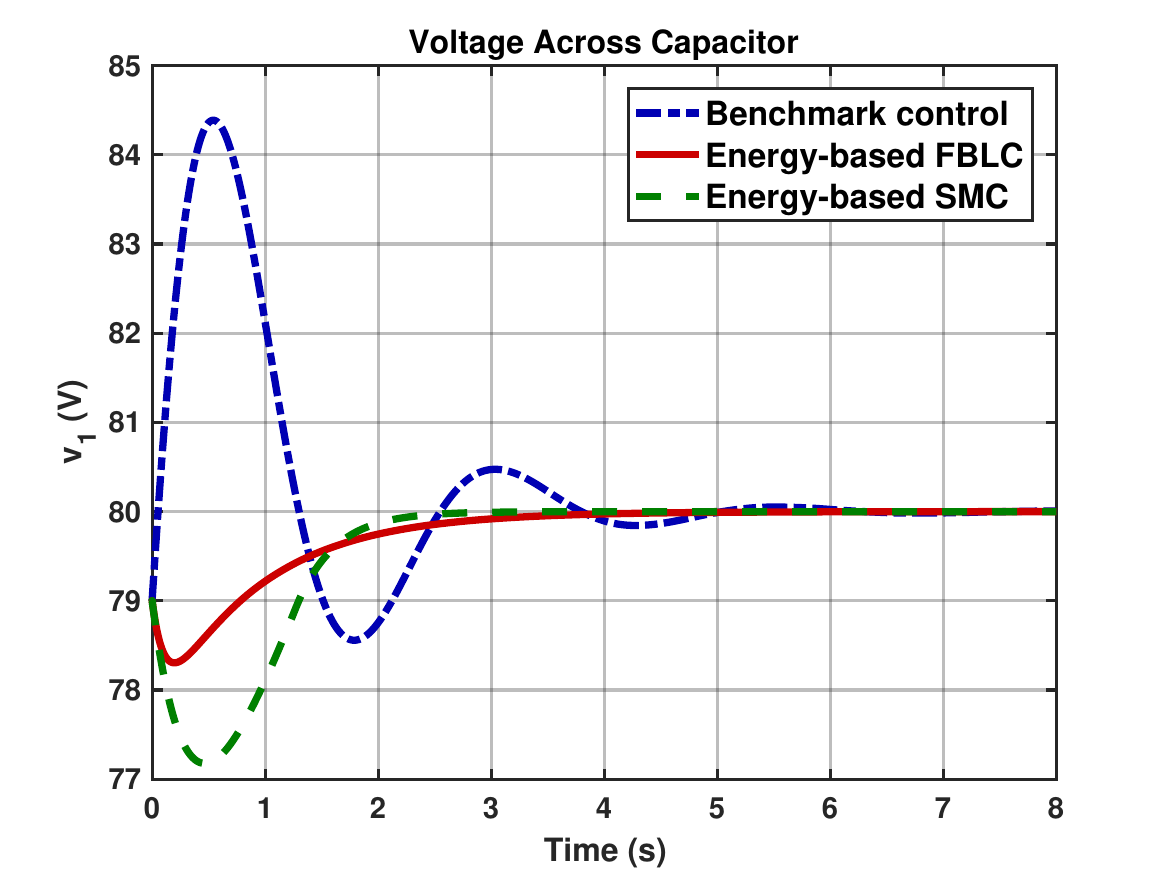}
        \caption{Voltage across capacitor}
        \label{fig:V1_tv}
    \end{subfigure}
    \hfill
    % Subfigure 3: Applied control voltage
    \begin{subfigure}[t]{0.48\textwidth}
        \centering
        \includegraphics[width=\textwidth]{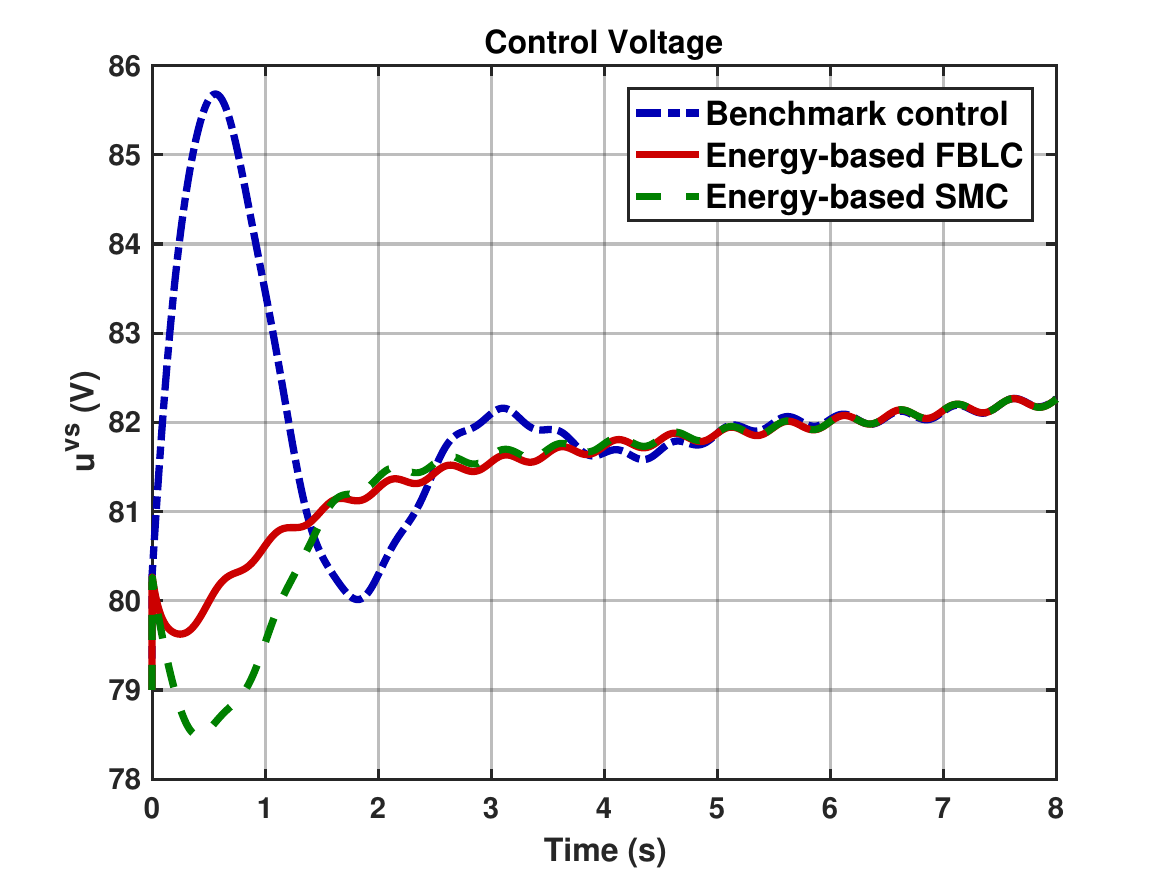}
        \caption{Applied control voltage}
        \label{fig:U_tv}
    \end{subfigure}
    \caption{System response of the RLC circuit under nonlinear benchmark control \eqref{eqn:brayton_moser}, energy-based FBLC, and energy-based SMC for a time-varying power load, with the objective of stabilizing and regulating the terminal voltage $v_1$ to $80~\text{V}$}
    \label{fig:IVUplot_tv}
\end{figure}

Consider the time-varying load profile as shown in Figure~\ref{fig:load_profile_tv}. 
Conventional constant-gain proportional controllers fail to regulate voltage under such time-varying power demands.
Therefore, we compare with the nonlinear Brayton–Moser-based controller in~\cite{cucuzzella2019voltage}, which guarantees stability for all $y^{ref}_1 > 0$ and $P_{load} \leq \Pi$,
\begin{align}
    u^{vs}_{bm} &= y_1^{ref} + R_1 i_1 - L_1\left( \frac{\Pi}{v_1^2} + N_3\right) \frac{dv_1}{dt} - \left( N_1 (v_1 - y_1^{ref}) + N_2 \frac{dv_1}{dt} \right) .\label{eqn:brayton_moser}
\end{align}
The controller gains are chosen as,
\begin{align}
    N_1 = 8, \quad N_2 = 1, \quad N_3 = 2, \quad \Pi = 3\cdot10^3 .\label{eqn:brayton_moser_gains}
\end{align}
Figure~\ref{fig:IVUplot_tv} shows that all controllers regulate voltage to $80~\text{V}$. 
The benchmark controller shows significant overshoot while 
the energy-based controllers give smoother transients and reduced control wear and tear.

\subsection{Frequency regulation in a synchronous generator}
\begin{figure}[ht]
    \centering
    \begin{subfigure}[t]{0.48\linewidth}
        \centering
        \includegraphics[width=\linewidth]{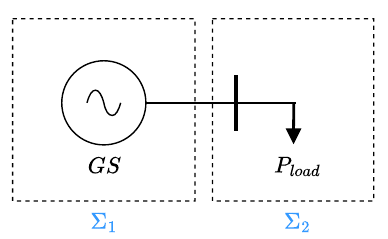}
        \caption{Generator connected to load}
        \label{fig:gs}
    \end{subfigure}
    \hfill
    \begin{subfigure}[t]{0.48\linewidth}
        \centering
        \includegraphics[width=\linewidth]{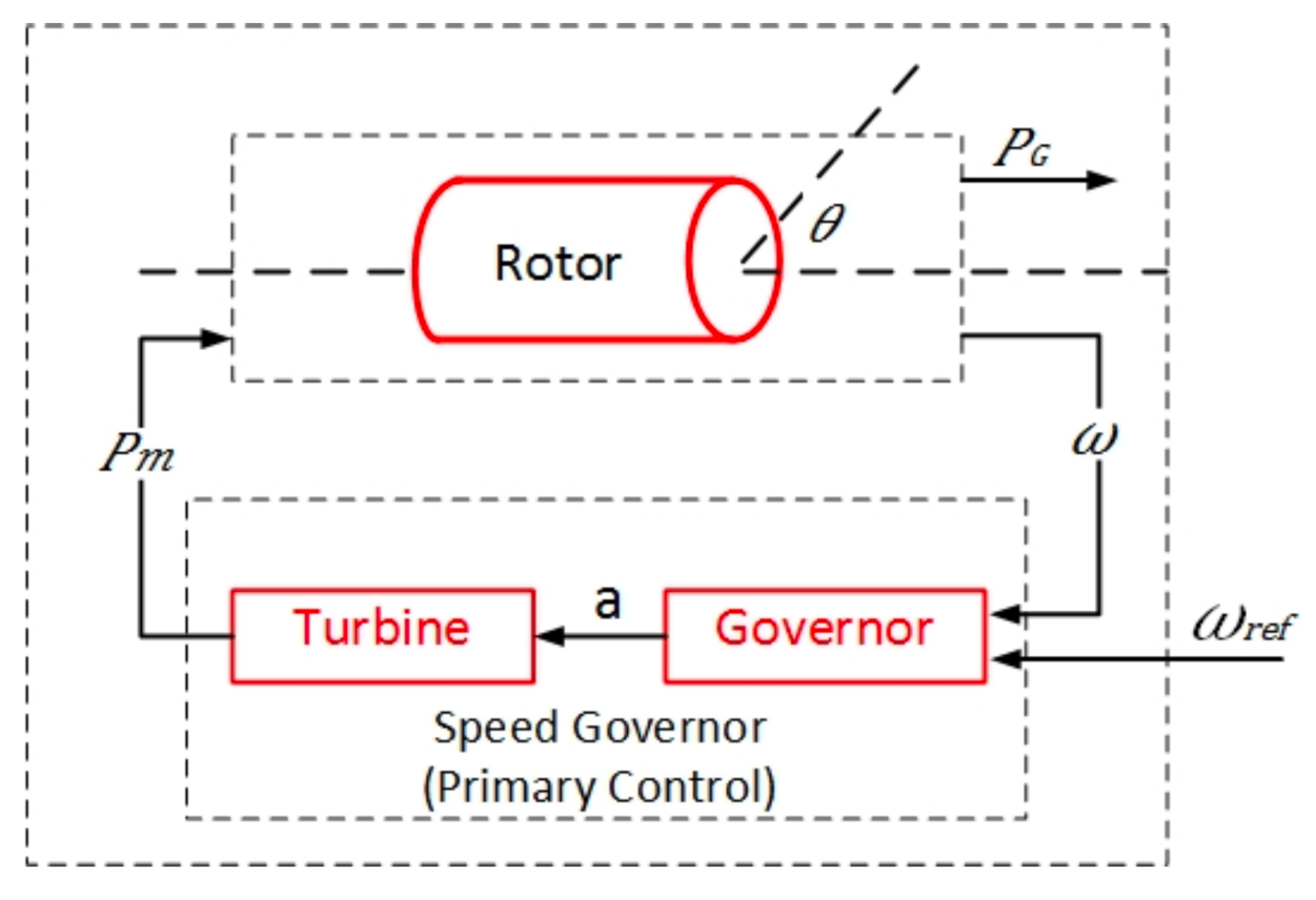}
        \caption{Generator–turbine–governor model}
        \label{fig:gtg}
    \end{subfigure}
    \caption{Synchronous generator with controllable turbine valve position $a_1$ ($\Sigma_1$) supplying power to a black-box load with unknown internal dynamics ($\Sigma_2$)}
    \label{fig:generator_system}
\end{figure}
Figure~\ref{fig:gs} depicts a synchronous generator supplying power to a load with specified power demand $P_{{load}}$ and
Figure~\ref{fig:gtg} shows a block diagram of the generator–turbine–governor system.
The objective is to regulate the generator frequency $\omega_1$ to $60$~Hz or $377$~rad/s under load variations.
The physical state, control and output variables of subsystem $\Sigma_1$ are,
\begin{align*}
    x_1 = \begin{bmatrix}
        \omega_1 \\ P_{m,1}
    \end{bmatrix}, \quad 
    u_1 = a_1, \quad y_1 = \omega_1,
\end{align*}
where $\omega_1$ is the rotor speed (generator frequency), $P_{m,1}$ is the turbine mechanical power, and $a_1$ is the governor valve position controlling the steam flow into the turbine.
For simplicity, the rotor is modeled by the classical swing equation and the turbine by an IEEE Type-1 linear model,
\begin{align*}
    \frac{d}{dt}\underbrace{\begin{bmatrix}
        \omega_1 \\ P_{m,1}
    \end{bmatrix}}_{x_1} &= \underbrace{\begin{bmatrix}
        -\frac{D_1}{J_1} & \frac{1}{J_1 \omega_1}\\
        0 & -\frac{1}{T_t} 
    \end{bmatrix} \begin{bmatrix}
        \omega_1 \\ P_{m,1}
    \end{bmatrix}}_{f_{x,1}(x_1)} + \underbrace{\begin{bmatrix}
        0 \\ \frac{K_t}{T_t}
    \end{bmatrix}}_{g_1^u(x_1)} \underbrace{a_1}_{u_1} + \underbrace{\begin{bmatrix}
        -\frac{1}{J_1 \omega_1} \\ 0
    \end{bmatrix}}_{g_1^r(x_1)} \underbrace{P_{load}}_{r_1},
\end{align*}
where $J_1$ is the moment of inertia, $D_1$ is the damping coefficient, $T_t$ is the turbine time constant, and $K_t$ is the turbine gain.
To design the energy-based control for $\Sigma_1$, we identify the relevant effort and flow variables,
\begin{align*}
    &
    \begin{aligned}
        e_1^u &= \frac{P_{m,1}}{\omega_1}, \quad &f_1^u &= \omega_1,
    \end{aligned}
     &&\hspace{-1cm} \text{(for } \Sigma_1 \text{)} \\
    &
    \begin{aligned}
        e_2^r &= \frac{P_{{load}}}{\omega_2}, \quad &f_2^r &= \omega_2.
    \end{aligned}
     &&\hspace{-1cm} \text{(for } \Sigma_2 \text{)} 
\end{align*}
From the definitions in Section~\ref{sec:energy_modeling}, $\Sigma_2$ computes $\zd_2^r$, with $P_2^r=P_{load}$ and $\Sigma_1$ computes $x_{z,1}$ using local state measurements.
The output of interest is the generator frequency, which is to be regulated to $y_1^{ref} = 377~\text{rad/s}$. 
The higher level energy space output references are thus chosen as,
\begin{align}\label{eq:eg2_ref}
    \mathbf{y}_{z,1}^{ref} &= \begin{bmatrix}
        E_1^{ref} \\ p_1^{ref}
    \end{bmatrix} = \begin{bmatrix}
        \frac{1}{2}J_1{y_1^{ref}}^2 
        \\0
    \end{bmatrix}.
\end{align}
Satisfaction of the conditions in Theorem~\ref{thm:interlayer_consistency} follows directly from~\eqref{eq:eg2_ref}.
For the synchronous generator subsystem, the mapping function $d_1^u(u_{z,1}, x_1)$ is characterized by the following relation,
\begin{align*}
    u_1 &= d_1^u(u_{z,i}, x_1) := \frac{T_t \left( 2 \dot{\omega}_1 \frac{P_{m,1}}{\omega_1} - u_{z,1} \right) + P_{m,1}}{K_t}.
\end{align*}
The simulation uses the following parameter values,
\begin{align*}
    J_1 = 10~\text{kg $\cdot$ m}^2, \; D_1 = 0.01~\text{N$\cdot$m$\cdot$s/rad}, 
    \; T_t = 0.5~\text{s}, \; K_t = 1000~\text{W/cm} .\nonumber
\end{align*}
with initial conditions, $\omega_1(0) = 373.23~\text{rad/s}$ and $P_{m,1}(0) = 1~\text{kW}$.

\begin{figure}[ht]
    \centering
    \begin{subfigure}[t]{0.48\textwidth}
        \centering
        \includegraphics[width=\textwidth]{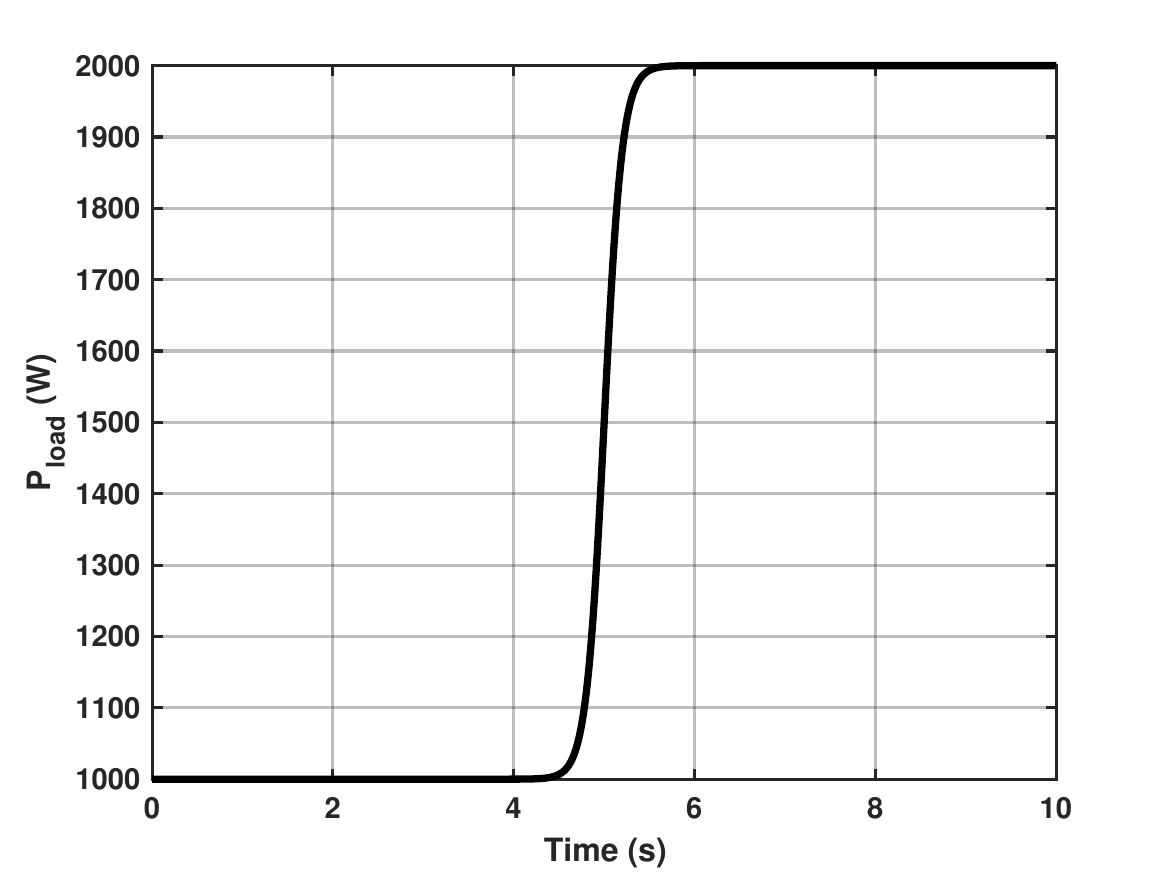}
        \caption{Smooth step (sigmoid) load profile}
        \label{fig:load_profile_sigmoid}
    \end{subfigure}
    \hfill
    \begin{subfigure}[t]{0.48\textwidth}
        \centering
        \includegraphics[width=\textwidth]{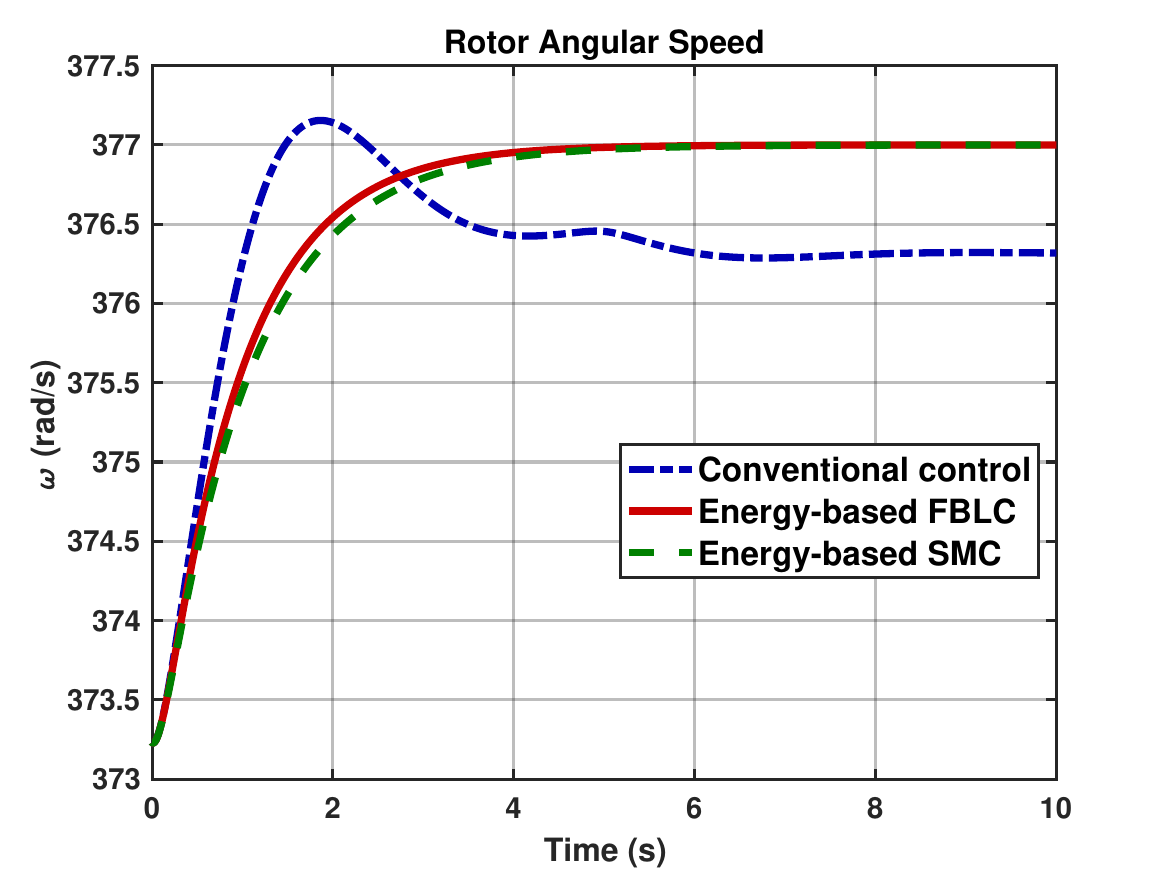}
        \caption{Rotor angular speed}
        \label{fig:o1_tv}
    \end{subfigure}
    \vspace{0.4em}
    \begin{subfigure}[t]{0.48\textwidth}
        \centering
        \includegraphics[width=\textwidth]{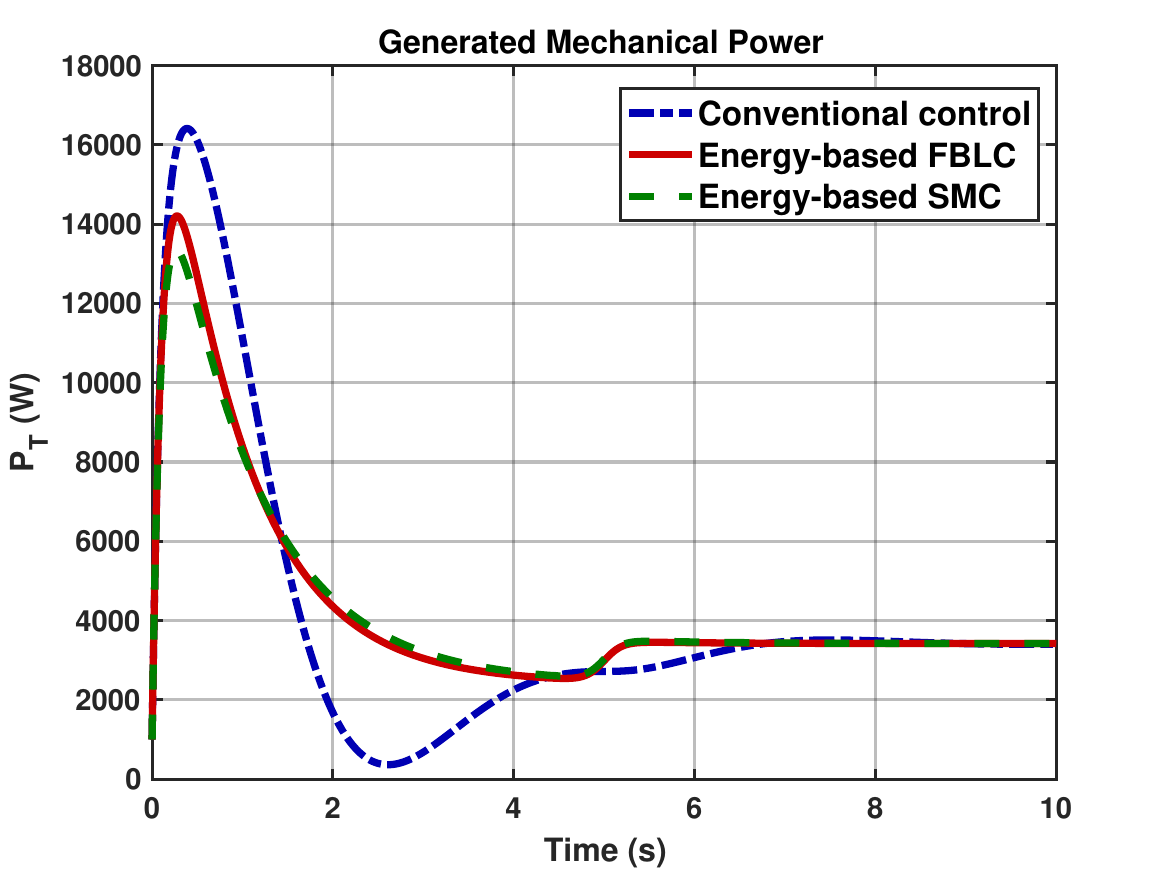}
        \caption{Generated mechanical power}
        \label{fig:Pm1_tv}
    \end{subfigure}
    \hfill
    \begin{subfigure}[t]{0.48\textwidth}
        \centering
        \includegraphics[width=\textwidth]{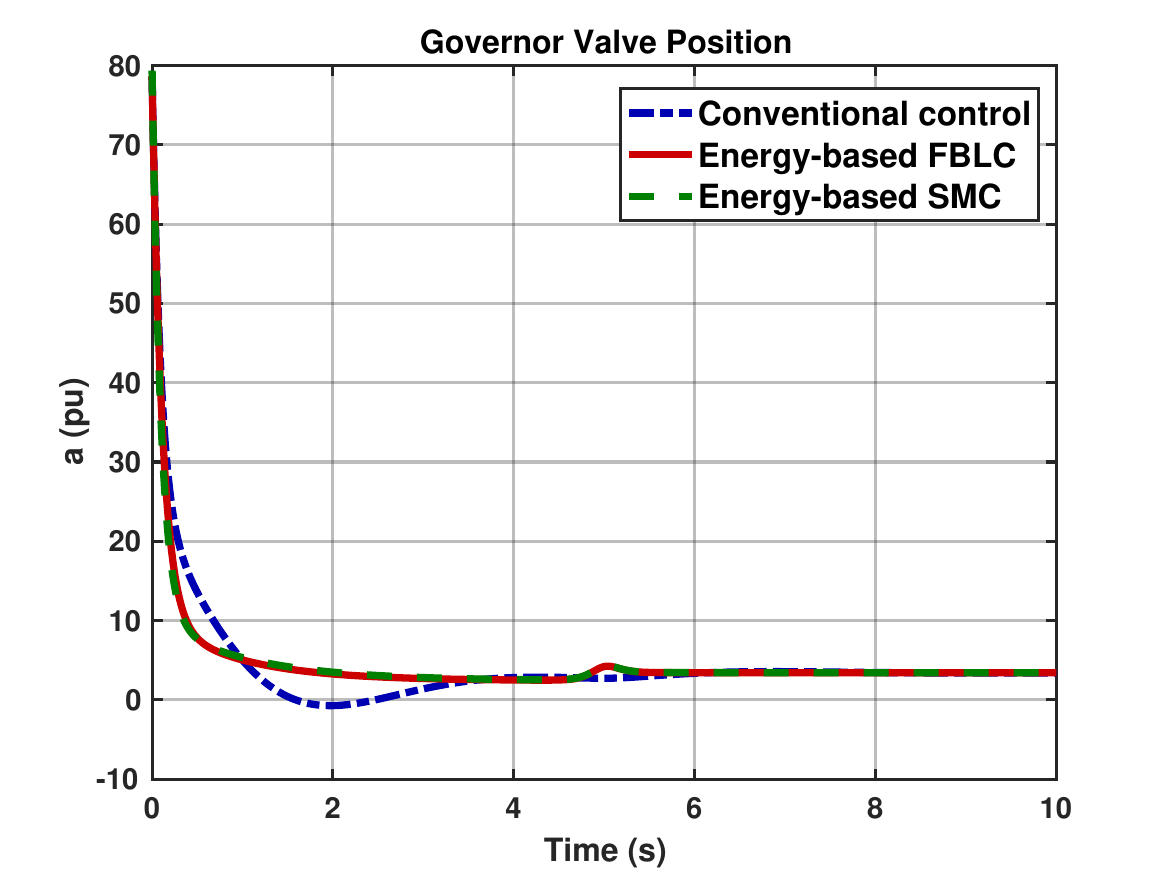}
        \caption{Governor valve position}
        \label{fig:a_tv}
    \end{subfigure}

    \caption{System response of the synchronous generator under conventional governor control~\eqref{eqn:valve_conv}, energy-based FBLC, and energy-based SMC for a smooth step load, with the objective of stabilizing and regulating the rotor angular speed (frequency) $\omega_1$ to $377~\text{rad/s}$}
    \label{fig:wPaplot}
\end{figure}

Consider a smooth step change in power demand, implemented using a sigmoid function as shown in Figure~\ref{fig:load_profile_sigmoid}.
As a benchmark, we simulate the conventional primary control scheme commonly used in practice,
\begin{align}\label{eqn:valve_conv}
    \dot{a}_1 = -\frac{a_1}{T_g} - \frac{1}{T_g}\frac{\omega_1 - y_1^{ref}}{r},
\end{align}
where $T_g$ is the governor time constant and $r$ is the droop coefficient.
For this example, the droop gain is chosen as $r = 0.2$.
The energy-based FBLC gains are selected as $K_1 = K_2 = 10$,
while the energy-based SMC method uses $M_\circ = 10$ and $M_1 = 1$.

Figure~\ref{fig:wPaplot} shows that the conventional droop controller in~\eqref{eqn:valve_conv} introduces steady-state error proportional to load change, whereas energy-based FBLC and SMC eliminate this error and improve transient response. 
Following the load variation around $t=5$~s, mechanical power adjusts faster under energy-based control. 
Overall, the proposed controllers outperform droop control in both transient and steady-state behavior.

\section{Conclusion}
\label{sec:conclusion}

This paper proposes a distributed component-level control framework using energy space modeling for interconnected power systems. 
A multilayered architecture is established in which component dynamics are lifted to an energy state space, where control is designed and mapped back to physical space for implementation without phasor approximations or PQ decoupling. 
An extended third-order energy space model is developed with tangent-space energy as a state to effectively capture higher-order dynamics. 
The resulting framework provides a technology-agnostic, modular, and scalable architecture requiring only local states and limited neighbor information. 
Distributed energy-based control is implemented on an RLC circuit and a synchronous generator, demonstrating improved voltage and frequency regulation, respectively, compared to conventional methods. 
Future work includes extending the framework to include system-level objectives and MIMO systems.

\bibliographystyle{unsrt}        
\bibliography{autosam}

\end{document}